\documentclass[]{aa}  
\usepackage{graphicx}
\usepackage{longtable}
\usepackage{lscape}
\usepackage{float}
\usepackage{subfig}
\usepackage{ulem}
\usepackage{txfonts}
\usepackage{natbib,array,tabularx,afterpage,booktabs}
\usepackage[figuresright]{rotating}
 \usepackage[hypertex,a4paper=true]{hyperref}
\usepackage{color}
\begin{document}

   \title{A low-luminosity type-1 QSO sample}
   \subtitle{III. Optical spectroscopic properties and activity classification} 

   \author{E.Tremou
          \inst{1,3}, M. Garcia-Marin\inst{1}, J. Zuther\inst{1}, A.
          Eckart\inst{1,2}, M. Valencia-Schneider\inst{1}, M. Vitale \inst{1,2}, C. Shan\inst{1}}  
   \institute{I.Physikalisches Institute, University of Cologne,
              Z\"ulpicher Strasse 77, 50937 Cologne, Germany\\
   \email{tremou@yonsei.ac.kr} 
              \and
             Max Planck Institute for Radioastronomy, Auf dem H\"ugel 69, 53121
             Bonn, Germany 
             \and
             Yonsei University Observatory, Yonsei University, Seoul 120-749,
             Republic of Korea} 
   \date{Received Month xx, 201x; accepted xxx x, 201x}

\abstract
{
We report on the optical spectroscopic analysis of a sample of 99 low-luminosity quasi-stellar objects (LLQSOs) at $z\leq 0.06$ base the
Hamburg/ESO QSO survey (HES). 
To better relate the low-redshift active galactic nucleus (AGN) to the QSO
population it is important to study samples of the latter type at a level of
detail similar to that of the low-redshift AGN.  
Powerful QSOs, however, are absent at low redshifts due to evolutionary effects and their small space density.
Our understanding of the (distant) QSO population is, therefore, significantly
limited by angular resolution and sensitivity. 
The LLQSOs presented here offer the possibility of studying the faint end of this
population at smaller cosmological distances and, therefore, in greater detail.
}
{In comparing two spectroscopic methods, we aim to establish a reliable activity classification scheme 
of the LLQSOs sample. 
Our goal is to enrich our systematic multiwavelength analysis of the AGN/starburst 
relation in these systems and give a  complementary information on this
particular sample of LLQSOs from the Hamburg ESO survey.
}
{
Here, we present results of the analysis of visible wavelength spectroscopy
provided by the HES and the 6  Degree Field Galaxy Survey (6dFGS). These surveys use different spectroscopic techniques, long-slit and circular fiber, respectively.
These allow us to assess the influence of different apertures on the
activity of the LLQSOs using classical optical diagnostic diagrams. 
We perform a Gaussian fitting of strong optical emission lines and decompose
narrow and broad Balmer components. 
}
{
A small number of our LLQSO present no broad component, which is likely to be present but buried in the noise. Two sources show double broad components, whereas six comply with the classic NLS1 requirements. As expected in NLR of broad line AGNs, the [S{\sc{ii}}]$-$based electron density values range between 100 and 1000 N$_{e}$/cm$^{3}$. Using the optical characteristics of Populations A and B, we find that 50\% of our sources with H$\beta$ broad emission are consistent with the radio-quiet sources definition. The remaining sources could be interpreted as low-luminosity radio-loud quasar. The BPT-based classification renders an AGN/Seyfert activity between 50 to 60\%. For the remaining sources, the possible starburst contribution might control the LINER and H{\sc{ii}} classification. Finally, we discuss the aperture effect as responsible for the differences found between data sets, although variability in the BLR could play a significant role as well. 
}
   {}

\titlerunning{LLQSO: spectroscopic properties and activity classification}
\authorrunning{E. Tremou et al.}
   \maketitle

\section{Introduction}
\label{sec:intro}
Accretion of matter onto a supermassive black hole (SMBH) at the center of a galaxy is the main energy source of galaxies hosting an active galactic nucleus (AGN). In a similar vein, 
the centers of starburst galaxies are not considered to be very active in terms of 
nonthermal emission arising from nuclear accretion and star formation processes originate the energy output instead. 
However, the association between the AGN activity and the star formation mechanism is still undetermined in galaxy evolution scenarios. Hence, reliable classification frames are vital to establish the activity of the galaxies.

Classification of AGN depends upon many parameters. Various studies
focusing on selection criteria, morphology, and line widths, have produced a
variety of classification schemes \citep[e.g.,][and references
therein]{2011MNRAS.413.1687C, 2008ARA&A..46..475H, 1995PASP..107..803U}. 
The emission line spectra of extragalactic sources has proven to be a reliable approach to diagnose 
the origin of the ionizing emission in a galaxy. 
In particular, the information contained in the relative
intensities of the emission lines in the visible
domain have been used by 
\citet[][BPT diagrams]{baldwin1981}, \cite{1987ApJS...63..295V}, and more recently by 
\cite{2006MNRAS.372..961K}.
The main idea is to discriminate between the different excitation mechanisms
operating on the line emitting gas. 
Depending on the contribution of the AGN, the galaxies can be categorized as
quasi-stellar objects (QSOs), from the high-power tail of the distribution in BPT diagrams  down
to Seyferts and low-ionization nuclear wmission line eegions
\citep[LINERs;][]{1980BAAS...12..809H}.

The nature of LINERs has been a long debate, with several explanations being offered to account for it. Ionization by shocks was one of the first, \citep{Heckman1980}, with young hot stars being responsible for it \citep{Terlevich1985, Dopita1995}. Pre-main sequence stars ionization \citep{Cid2004} have also been proposed as ionization sources. Ionization by low-luminosity AGNs is a favored explanation \citep{Ferland1983,Halpern1983,Ho1997}, in which case they would constitute the main fraction of the AGN population. More recently, using radial emission-line surface brightness profiles, \citet{Singh2013} found that the class of LINER galaxies are not generally uniquely powered by a central AGN. They postulate that the excess LINER-like emission is ionization by evolved stars during the short, but very hot and energetic phase known as post-AGB.

Starburst galaxies are mainly ionized by hot stars
\citep{1977ApJ...217..928H,1992ApJ...397L..79F,1992ApJ...399L..27S,
1998AJ....116...55M,2000PASP..112..753B}. One main feature of the optical spectra in these sources is the presence of emission lines.
Narrow permitted and forbidden emission lines (300-1000 km/s width) originating
in the narrow line region (NLR) and especially broad lines (2000-6000 km/s
width) originating in the broad line region (BLR) are considered to be an
unambiguous indication of an AGN \citep{oster1987}. Some objects, such as QSOs and
Seyfert 1 galaxies, show both types of lines. Seyfert 2s and LINERs show only
narrow line emission. 

Activity classification and diagnostic schemes relying on optical spectroscopic observations need to be handled carefully. The extended emission of the host galaxy may contaminate the point source nuclear spectral light. 
Early studies  discussed the observational effects on early-type galaxies, which appear to be redder in their centers \citep{1961ApJS....5..233D,1963AJ.....68..237H} . The color-aperture relation in early-type spiral galaxies has been shown to depend on the redshift and the size of the aperture \citep{1971Ap&SS..12..394T}.
The H$\alpha$ emission line is widely used for estimations of star formation rates (SFR, e.g., \citet{Kennicutt1983}) and more recent analyses have dealt with the impact of aperture, which can lead to substantial fractions of the emission-line flux loss \citep{1995AJ....110.1602Z,2003ApJ...591..827P}. Furthermore, \cite{2003ApJ...599..971H}  concluded that the H$\alpha$ SFRs of SDSS samples are overestimated for nearby or most massive galaxies. The comparison with the radio SFRs  led to a large deviation of the H$\alpha$ SFRs  and the quantification of aperture effect was challenging. 
\cite{2005PASP..117..227K} investigated the effect of aperture on metalicity, extinction, and star formation rate by studying the integrated and nuclear spectra of a sample of galaxies as a function of both galaxy type and luminosity. They found that for flux covering fractions $<$ 20\% of the galaxy light, the difference between the nuclear and global metalicity, extinction and star formation rate is substantial.

The influence of the atmospheric seeing effects on measurements of the [O\,{\sc{iii}}]/H$\beta$ flux ratio and, in particular, for the case of the NLR of Seyfert galaxies is discussed in a study by \cite{1983ApJ...270...71P}, using models of the surface brightness distribution. Their model resulted in a systematic effect of the [O\,{\sc{iii}}]/H$\beta$ flux ratio at the 25\% level for apertures close to the size of the NLR (usual case for the near Seyfert 1 galaxies).
\cite{1992A&A...266...72W}  show the importance of the seeing effects in variability studies of low-luminosity Seyfert 1 galaxies by simulating the seeing variations. Since the BLR and the continuum source are not resolved, the seeing can drive the uncertainties on the measured ratio of the narrow line flux to BLR. 
 
Recently, \citet{2014MNRAS.441.2296M}  calculated the mean dispersion for the diagnostic line ratios used in the standard BPT diagram with respect to the central aperture of central extraction to obtain an estimate of the uncertainties resulting from aperture effects. They found that the starlight subtraction does not significantly change  the effect of a different placement in the BPT diagrams, which results from a fiber and standard slit observational methods.

The main purpose of this paper is to characterize a sample of LLQSOs via the analysis of their optical spectra. We aim to classify their nuclear activity and to provide a more in-depth study of the aperture effect using a unique sample of low-luminosity AGN in the local universe. The sample members are among the closest AGN and have been studied at high angular resolution and with
a variety of methods \citep{2006A&A...452..827F,bertram2007,konig2009,Busch2013B, Busch2015C, Busch2015D, Moser2015}.
In light of the importance and rarity of this sample, a complete and robust characterization of these galaxies is a crucial first step toward securing its full scientific return in terms of future studies with current and next generation observing facilities (e.g., the Atacama large millimeter array - ALMA, the square kilometer arry - SKA, the James Webb telescope - JWST).

This paper is structured as follows. In Section \ref{sample}, we introduce the low-luminosity QSO sample and its selection criteria. The data we used for this study are presented in Section \ref{obser}. The spectroscopic analysis that we followed is described in Section \ref{anal}. Section \ref{fitres} presents the fitting results and draws some general sample characteristics based on them. The activity classification scheme of the sample is shown in Section \ref{class}, also comparing  the results obtained using different observational techniques. Moreover, we compare the activity classification for both fiber and long-slit methods. Finally, we discuss the impact of the aperture effect in Section \ref{discussion} and  summarize in Section \ref{concl}.

\section{The low-luminosity QSO sample}\label{sample}

Our low-luminosity QSO sample is drawn from the Hamburg/ESO QSO survey
\citep[hereafter HES;][]{1996A&AS..115..227W,reimers1996,wisotzki2000}, 
a wide angle ($\approx$9500 deg$^2$) Southern Hemisphere survey for optically bright QSOs. The HES selects Type 1 AGNs up to a redshift of about 3.2, and has a brightness limit of $B_{J} \lesssim 17.3$. Because of variations on the observed field, this brightness limit has a dispersion of $\sim$ 0.5 mag.

Classic QSO detection techniques \citep[e.g.,][]{1983ApJ...269..352S} tend to miss luminous AGN, especially at low redshift. This introduces a distance-dependent incompleteness, which is not uncommon in QSOs surveys. Comparatively, the method used by the HES facilitates the inclusion of bright extended objects, which are located at the faint end of the distribution (17 $\lesssim B_{J} \lesssim$18).
We selected a subsample of 99 objects from the HES (see Table \ref{tab:LLQSOSample}). The main selection criteria was that they should all have a small cosmological distance, $z \leq 0.06$, thus ensuring  the presence of the stellar CO(2-0) band head in the near-infrared 
(NIR) $K$-band (e.g., \citealt{2006A&A...452..827F}).

\begin{figure}[h]
\centering
\includegraphics[width=\columnwidth]{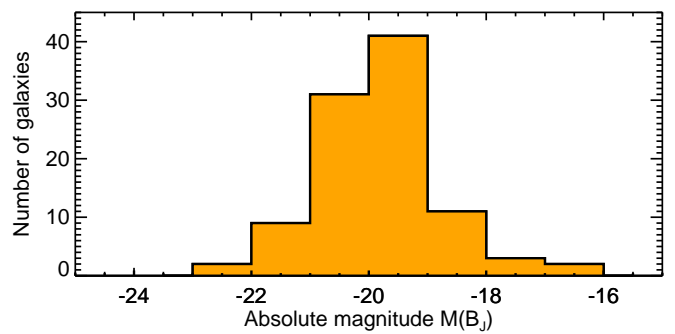}
\caption{Absolute $B_J$ magnitude distribution of the LLQSOs sample.}
\label{fig:magn}
\end{figure}

Figure \ref{fig:magn} presents the absolute $B_J$ magnitude distribution of
the sample. Absolute magnitudes, $M(B_J)$, were estimated from the
foreground extinction corrected apparent magnitudes and assuming an AGN
power-law, $F_\nu\propto \nu^{-\alpha}$ with index $\alpha=0.5$
\citep{2001AJ....122..549V}.
The absolute magnitude covers the range $-23\lesssim M(B_J)\lesssim -16$.

\begin{figure}[h]
\begin{center}
\resizebox{\hsize}{!}{\includegraphics[width=0.5\textwidth]{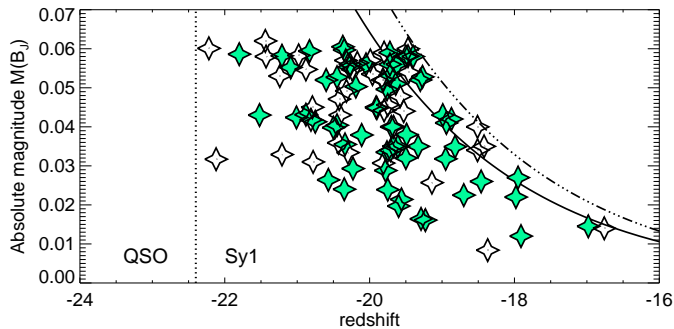}}
\end{center}
\caption{Redshift magnitude diagram of the sample. All sources  observed by the HES. Filled symbols represent thoes observed by the 6df as well. The dotted vertical line marks the classical demarcation between QSO and Seyfert 1 galaxies. The solid line represents the magnitude limit of the HES ($B_{J} < 17.3$), whereas the dot-dashed line marks the 0.5 mag dispersion.}
\label{fig:redshift}
\end{figure}

As shown in Fig\,\ref{fig:redshift}, following the commonly used demarcation between QSOs and Seyfert1 $M_{B} =  −21.5 + 5log h_{0}$ our subsample falls into the Seyfert 1 region, slightly under luminous to be classified as QSOs. We then chose to name the objects presented here as low-luminosity QSOs (LLQSOs).

\onecolumn
\begin{longtable}{cllccl}
\caption{Main characteristics of the LLQSOs sample: HES and 6dFGS names, J2000 coordinates and redshift. $\ddagger$ marks galaxies for which we do not have HES optical spectra. \label{tab:LLQSOSample}}\\
\hline
\hline
\rule[-1ex]{0pt}{1.5ex} {ID} & {HES name} &{6dFGS name} & {RA [deg]} & {DEC [deg]}  & {redshift}\\ 
\hline
\endfirsthead
\caption{continued}\\
\hline\hline
\rule[-1ex]{0pt}{1.5ex} {ID} & {HES name} &{6dFGS name} & {RA [deg]} & {DEC [deg]}  & {redshift}\\
\hline
\endhead
\hline
\endfoot
1  & HE0003-5023 &                &  1.42917     & -50.11530  &0.0334\\ 
2  & HE0021-1810 &                &  5.91417     & -17.89810  &0.0535\\
3  & HE0021-1819 &g0023554-180251 &  5.98042     & -18.04720  &0.0532\\
4  & HE0022-4546{$^\ddagger$} &g0025013-452955 &  6.25500     & -45.49830  &0.056\\
5  & HE0036-5133{$^\ddagger$} &g0039159-511702 &  9.81583     & -51.28390  &0.0288\\
6  & HE0038-0758 &                &  10.21960    & -7.70278   &0.054\\
7  & HE0040-1105 &g0042369-104922 &  10.65330    & -10.82250  &0.042\\
8  & HE0045-2145 &g0047413-212927 &  11.92210    & -21.49080  &0.0214\\
9  & HE0051-2420 &g0053544-240437 &  13.47670    & -24.07670  &0.056\\
10 & HE0103-3447 &                &  16.44420    & -34.52920  &0.057\\
11 & HE0103-5842 &                &  16.32080    & -58.43780  &0.0257\\
12 & HE0108-1631 &g0111143-161555 &  17.80920    & -16.26500  &0.052\\
13 & HE0108-4743{$^\ddagger$} &g0111097-472735 &  17.79040    & -47.46000  &0.0239\\
14 & HE0111-1506{$^\ddagger$} &g0113499-145057 &  18.45750    & -14.84920  &0.0527\\
15 & HE0114-0015 &                &  19.26500    &  0.00750   &0.0456\\
16 & HE0119-0118 &                &  20.49920    & -1.04028   &0.0547\\
17 & HE0122-5137 &                &  21.25080    & -51.36580  &0.052\\
18 & HE0125-1904 &                &  22.02790    & -18.80860  &0.043\\
19 & HE0126-0753 &g0129067-073830 &  22.27750    & -7.64167   &0.056\\ 
20 & HE0149-3626 &g0151419-361116 &  27.92460    & -36.18780  &0.0335\\
21 & HE0150-0344 &                &  28.25580    & -3.49000   &0.0478\\
22 & HE0203-0031 &g0206160-001729 &  31.56620    & -0.29139   &0.0424\\
23 & HE0212-0059 &g0214336-004600 &  33.64000    & -0.76667   &0.0264\\
24 & HE0224-2834 &g0226257-282059 &  36.60710    & -28.34970  &0.0605\\
25 & HE0227-0913{$^\ddagger$} &g0230055-085953 &  37.52250    & -8.99806   &0.0164\\
26 & HE0232-0900 &g0234378-084716 &  38.65710    & -8.78778   &0.043\\
27 & HE0236-3101 &                &  39.68790    & -30.80670  &0.062\\
28 & HE0236-5224 &                &  39.58210    & -52.19220  &0.045\\
29 & HE0253-1641 &g0256027-162916 &  44.01080    & -16.48780  &0.032\\
30 & HE0257-2434 &g0259305-242254 &  44.87710    & -24.38170  &0.035\\
31 & HE0323-4204 &                &  51.25920    & -41.90500  &0.058\\
32 & HE0330-1404 &g0333078-135433 &  53.28210    & -13.90940  &0.04\\
33 & HE0332-1523 &g0334245-151340 &  53.60210    & -15.22810  &0.035\\
34 & HE0336-5545 &                &  54.52620    & -55.60000  &0.059\\
35 & HE0342-2657{$^\ddagger$} &g0345032-264820 &  56.26330    & -26.80530  &0.058\\
36 & HE0343-3943{$^\ddagger$} &g0345125-393429 &  56.30210    & -39.57500  &0.0431\\
37 & HE0345+0056 &                &  56.91750    & 1.08722    &0.031\\
38 & HE0349-4036 &g0351417-402759 &  57.92330    & -40.46640  &0.0582\\
39 & HE0351+0240 &                &  58.53920    & 2.82500    &0.034\\
40 & HE0358-3713 &g0400407-370506 &  60.16960    & -37.08530  &0.051\\
41 & HE0359-3841 &g0401462-383320 &  60.44250    & -38.55580  &0.059\\
42 & HE0403-3719 &g0405017-371115 &  61.25670    & -37.18780  &0.0552\\
43 & HE0412-0803 &g0414527-075540 &  63.71920    & -7.92806   &0.0379\\
44 & HE0429-0247 &g0431371-024124 &  67.90420    & -2.69028   &0.041\\
45 & HE0429-5343 &                &  67.66670    & -53.61560  &0.04\\
46 & HE0433-1028 &g0436223-102234 &  69.09250    & -10.37580  &0.0355\\
47 & HE0433-1150\footnote{Spectral resolution about 770\,km/s.} &                 &  68.88540    & -11.74030  &0.058\\
48 & HE0436-4717 &                &  69.36710    & -47.19140  &0.053\\
49 & HE0439-0832 &                &  70.47500    & -8.44306   &0.045\\
50 & HE0444-0513 &                &  71.83580    & -5.13722   &0.0442\\
51 & HE0447-0404 &g0450251-035903 &  72.60420    & -3.98389   &0.022\\
52 & HE0521-3630{$^\ddagger$} &g0522580-362731 &  80.74170    & -36.45890  &0.0553\\
53 & HE0535-4224 &g0537331-422230 &  84.38750    & -42.37500  &0.035\\
54 & HE0608-5606 &g0609175-560658 &  92.32330    & -56.11610  &0.0318\\
55 & HE0853-0126 &g0856178-013807 &  134.07401   &  -1.63528  &0.0597\\
56 & HE0853+0102 &                &  133.97600   &  0.85278   &0.052\\
57 & HE0934+0119\footnote{Spectral resolution about 880\,km/s.} &                 &  144.25400   &  1.09528   &0.0503\\
58 & HE0949-0122 &g0952191-013644 &  148.07899   &  -1.61222  &0.0197\\
59 & HE1011-0403 &g1014207-041841 &  153.58600   &  -4.31139  &0.0586\\
60 & HE1013-1947 &                &  153.98500   &  -20.04080 &0.0547\\
61 & HE1017-0305 &                &  154.88699   &  -3.33750  &0.0492\\
62 & HE1029-1831\footnote{Spectral resolution about 880\,km/s.} &g1031573-184633 &  157.98900   &  -18.77610 &0.0404\\
63 & HE1107-0813 &                &  167.45200   &  -8.50417  &0.0583\\
64 & HE1108-2813 &g1110480-283004 &  167.70000   &  -28.50080 &0.024\\
65 & HE1126-0407 &                &  172.31900   &  -4.40222  &0.0601\\
66 & HE1136-2304 &g1138510-232135 &  174.71300   &  -23.36000 &0.027\\
67 & HE1143-1810 &                &  176.41901   &  -18.45470 &0.0329\\
68 & HE1237-0504{$^\ddagger$} &           &  189.91400   &  -5.34444  &0.0084\\
69 & HE1248-1356 &g1251324-141316 &  192.88499   &  -14.22140 &0.0145\\
70 & HE1256-1805{$^\ddagger$} &           &  194.67900   &  -18.36000 &0.014\\
71 & HE1310-1051 &g1313058-110742 &  198.27400   &  -11.12830 &0.034\\
72 & HE1319-3048 &g1321582-310426 &  200.49200   &  -31.07360 &0.0448\\
73 & HE1328-2508 &g1331138-252410 &  202.80800   &  -25.40280 &0.026\\
74 & HE1330-1013 &g1332391-102853 &  203.16299   &  -10.48140 &0.0225\\
75 & HE1338-1423 &                &  205.30400   &  -14.64440 &0.0418\\
76 & HE1346-3003 &g1349193-301834 &  207.33000   &  -30.30970 &0.0161\\
77 & HE1348-1758 &g1351295-181347 &  207.87300   &  -18.22970 &0.012\\
78 & HE1353-1917 &g1356367-193145 &  209.15300   &  -19.52890 &0.0349\\
79 & HE1417-0909 &                &  215.02600   &  -9.38694  &0.044\\
80 & HE2112-5926 &                &  318.96500   &  -59.23170 &0.0317\\
81 & HE2128-0221 &g2130499-020814 &  322.70801   &  -2.13750  &0.0528\\      
82 & HE2129-3356 &g2132022-334254 &  323.00900   &  -33.71500 &0.0293\\
83 & HE2204-3249 &g2207450-323502 &  331.93799   &  -32.58390 &0.0594\\
84 & HE2211-3903{$^\ddagger$} &g2214420-384823 &  333.67499   &  -38.80670 &0.0398\\
85 & HE2221-0221{$^\ddagger$} &           &  335.95700   &  -2.10361  &0.057\\
86 & HE2222-0026 &g2224353-001104 &  336.14700   &  -0.18444  &0.0581\\
87 & HE2231-3722 &g2234409-370644 &  338.67099   &  -37.11220 &0.043\\
88 & HE2233+0124 &                &  338.92499   &  1.65917   &0.0564\\
89 & HE2236-3621 &                &  339.77200   &  -36.09810 &0.06\\
90 & HE2251-3316 &g2253587-330014 &  343.49399   &  -33.00360 &0.056\\
91 & HE2254-3712 &g2257390-365607 &  344.41199   &  -36.93530 &0.038\\
92 & HE2301-3517 &                &  346.15500   &  -35.02000 &0.04\\
93 & HE2302-0857 &                &  346.18100   &  -8.68583  &0.0471\\
94 & HE2306-3246 &g2309192-322958 &  347.32999   &  -32.49940 &0.052\\
95 & HE2322-3843 &                &  351.35101   &  -38.44690 &0.0359\\
96 & HE2323-6122 &g2326376-610602 &  351.65701   &  -61.10030 &0.0413\\
97 & HE2337-2649 &g2340321-263319 &  355.13300   &  -26.55530 &0.0496\\
98 & HE2343-5235 &                &  356.43500   &  -52.31000 &0.035\\ 
99 & HE2354-3044 &                &  359.36700   &  -30.46110 &0.0307\\
\hline 
\end{longtable}
\twocolumn

The LLQSO sample has been studied intensively at multiple wavelengths with both photometry and spectroscopy.
For all sample members accessible from the northern hemisphere, we
obtained millimeter measurements of CO(1-0) and CO(2-1) \citep[39 out of 99
objects][]{bertram2007}. For 27 of the CO detected objects, we  also obtained
21~cm HI measurements \citep{konig2009}. Both observations show that LLQSO hosts
are rich in cold molecular/atomic gas ($M_{\rm H_2}\approx 5\times 10^9
M_\odot$, $M_{\rm HI}\approx 10^{10}M_\odot$) and that the molecular gas is
concentrated in the central kpc (\citealt{2007A&A...464..187K}; \citealt{Moser2012}). 
In the near-infrared (NIR), so far we studied nine LLQSOs with the VLT. For eight of those, we
obtained ISAAC long-slit $K$-band spectra \citep{2006A&A...452..827F} and for the
ninth, we obtained adaptive-optics-assisted $H+K$ imaging spectroscopy with
SINFONI \citep{fischer_lrs_2008}. 
The current NIR data reveal predominantly late-type hosts with a high
incidence of bars. The AGN-subtracted colors in the NIR are typical for
nonactive, late-type galaxies. 

More recently, \citet{Busch2013B}  studied a subsample of 20 of our LLQSOs, performing aperture photometry 
and a decomposition into bulge, disk, bar, and bar components in the NIR. 
In good agreement with \citet{fischer_lrs_2008}, the analysis reveals that 50\% 
of hosts are disk galaxies, 86\% of them barred. The study also reveals stellar and black 
hole masses lower than those typical for brighter QSOs. 
Also, these LLQSOs do not follow the M$_{BH}$-L$_{bulge}$ relation for inactive galaxies. 

\section{Data used}\label{obser}
\begin{figure}[h]
\begin{center}
\resizebox{\hsize}{!}{\includegraphics[width=0.5\textwidth]{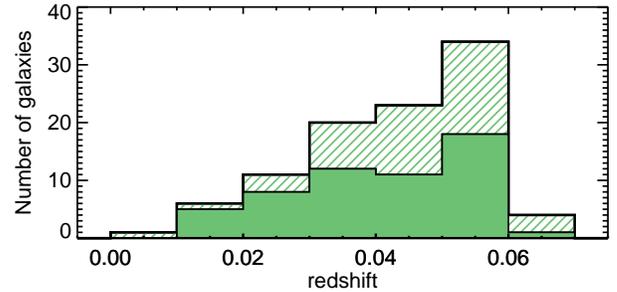}}
\end{center}
\caption{Redshift distribution of the LLQSO sample (as given in Tables\,\ref{tab:LLQSOSample}). Sources from 6dFGS  are represented with a solid histogram, while HES sources with  dashed histogram.}
\label{fig:hist_redshift}
\end{figure}

The optical spectroscopic analysis of the LLQSO sample is based on two sets of observations. 
The first data set is from the HES \citep{reimers1996}, in which follow-up observations were carried out to  confirm  the type-1 character of the survey candidates spectroscopically. 
In addition to this, we have used data from the 6 Degree Field Galaxy survey (6dFGS), an optical spectroscopy public database \citep{jones2004, jones2009}. The 6df data generally offers a better spectral resolution, and combined with the HES allows us to discuss aspects like aperture effect and its impact in the results. The redshift distribution of both data sets is very comparable (see Fig.\,\ref{fig:hist_redshift}), with no bias of the 6dFGS subsample toward lower or higher redshifts. The HES provides 71 and the 6dFGS 58 spectra of our 99 LLQSO sample (see details in Table\,\ref{tab:LLQSOSample}).

\subsection{The HES data}
The HES used three ESO telescopes (3.6~m,
2.2~m, 1.52~m) to obtain long-slit, low-resolution spectra ($R\approx 
165- 938$, i.e ${\rm FWHM}\approx 1800-320\,{\rm km}\,{\rm s}^{-1}$) during five
observing campaigns between 1990 and 1994. 
Depending on the seeing conditions, the slit width varied between $1\farcs 5$
and $2\farcs 5$, while the slit position angle was always east-west. 
A detailed description of the instrument setup and data reduction is given
in \cite{reimers1996}.  

We optimally extracted spectra  with an algorithm similar to that of \citet{1986PASP...98..609H}. 
Leaving only the amplitudes as free parameters, the extraction
procedure minimizes contamination from nearby sources, while at the same time
the extraction windows (and subsequent photometry) are limited to apertures 
the size of the seeing disk. We applied corrections for atmospheric extinction  using a standard La Silla extinction curve \citep{1995Msngr..80...34B}. The
flux calibration accuracy, monitored by standard star spectra, is in many cases
better than $\lesssim 20$\%, but because of instrumental limitations, data should not be considered spectrophotometric \citep{reimers1996}
However, since we are using flux ratios, the lack of spectrophotometrically calibrated data does not affect the results.

\subsection{The archive 6dFGS data}
The 6dFGS is designed to measure the redshift and peculiar velocities of a NIR sample of galaxies selected from the Two Micron All Sky Survey (2MASS). In addition, a variety of targets from other surveys were included as filler programs. A detailed description of the survey can be found in \cite{jones2004}. 
The instrument used to carry out the 6dFGS was the Six-Degree Field multiobject fiber spectrograph facility (6dF, \citealt{Parker1998}), located at the 1.2~m Anglo-Australian Observatory's UK Schmidt Telescope (UKST). It is able to record 150 simultaneous spectra over a 5.7~deg field. Spectra are obtained with a $6\farcs7$ diameter fiber, using two separate V and R gratings, which together give $R\approx 1000$ (corresp. to about ${\rm FWHM}\approx 300\,{\rm km}\,{\rm s}^{-1}$)
and cover at least 4000- 7500 \AA{.} Typical signal-to-noise ratios are $\sim 10$ per pixel. 
The same fixed average spectral transfer function is assumed for each plate all the time. Differences in the transfer function between individual 
fibers are
corrected for by the flat-fielding. All data are flux calibrated using 6dF observations of the standard stars Feige110 and EG274.
See \cite{jones2004, jones2009} for full details on the spectra processing.
We retrieved the already reduced data for our cross-matched sample from the publicly accessible online
database\footnote{\url{http://www-wfau.roe.ac.uk/6dFGS/}}. We make use of 6dFGS data release 3 (DR3), which is the final redshift release.

\section{Spectral line fitting} \label{anal}
The analysis of the optical spectra for both HES and 6dFGS focuses on seven emission lines:
H$\alpha$ $\lambda$ 6562\AA{} (hereafter H$\alpha$), 
H$\beta$ $\lambda$ 4861\AA{} (hereafter H$\beta$), 
[N\,{\sc{ii}]} $\lambda$$\lambda$ 6548, 6583\AA{}, 
[O\,{\sc{iii}]} $\lambda$$\lambda$ 4959, 5007\AA{}, 
[S\,{\sc{ii}}] $\lambda$$\lambda$ 6717,6731\AA{}, and
[O\,{\sc{i}}] $\lambda$ 6300\AA{}. These lines were selected to classify the host galaxies using the standard BPT diagnostic diagrams \citep{baldwin1981}.

The basic assumption here is that, since the hosts in the HES sources are usually relatively faint, the stellar contribution to the Balmer-line measurements is, in general, negligible.

There are, however, several cases in which Fe{\sc{ii}} and MgIb 5180 features are clearly seen, so we have assessed their impact in the fitting results. It is not uncommon for Seyfert1 galaxies to show broad and Fe{\sc{ii}} emission lines features in the blue region of the optical spectra \citep{Dong2011}. This emission surrounds and potentially contaminates the H$\beta$+[O{\sc{iii}}] complex, although its presence and strength is not directly related to the lines under study.  For the 6dFGS galaxies in our sample, 20\% exhibit clear Fe{\sc{ii}} features, whereas 24\% show unclear or low-level traces of it. In our HES data, these values are 30\% and 53\%, respectively. We  estimate that, in galaxies with a strong Fe{\sc{ii}} contribution, this may lead to a maximum uncertainty of 15\% in the emission lines flux measurements of the H$\beta$ and [O{\sc{i}}] lines. Depending on the case, this would add up an uncertainty of up to 0.13 dex on the O[{\sc{iii}}]/H$\beta$ axis of the BPT diagnostic diagrams.

Old stellar populations are responsible for the MgIb absorption feature that can bias the narrow lines fitting results \citep{Bica1986}. In this case 25\% of our 6dFGS and 27\% of our HES source clearly exhibit it, whereas 22\% and 17\% of the studies 6dFGS and HES spectra, respectively, present it at a low level. In this case, we do not aim to study the galaxies' stellar populations; we hence chose to fit the nebular emission lines using restricted spectral windows, and locally fit the continuum in each case with a first order polynomial fit. When broad absorption features, such as that of MgIb, are present, they would be accounted for as part of the continuum. This approach is not far from the standard SDSS estimate of the stellar continuum using a sliding median, which has been proven to be adequate for strong emission lines. This may be problematic if the aim is to recover weaker stellar or Balmer absorption lines \citep{Tremonti2004}, but they are not the focus of the present paper.

For each spectrum, the total integrated flux, central wavelength, and width of the relevant emission lines have been measured fitting Gaussian functions using the MPFIT IDL software package \footnote{\url{http://www.physics.wisc.edu/~craigm/idl/fitting.html}} \citep{Markwardt2009}. We fit all lines  using narrow components, except for H$\alpha$ and H$\beta$ , which  present both narrow and broad (coming from the narrow and broad line regions, respectively).  It is well known that, under certain conditions such as line blending, the fitting algorithms may derive different solutions. Some of our HES spectra suffer from line blending, especially in the H$\alpha$+[N\,{\sc{ii}}] complex and [S\,{\sc{ii}}] spectral regions. Therefore we decided to apply some restrictions to increase the reliability of our results.

\begin{figure*}
\centering \subfloat[]{\label{fig:allhb}\includegraphics[width=0.5\textwidth]{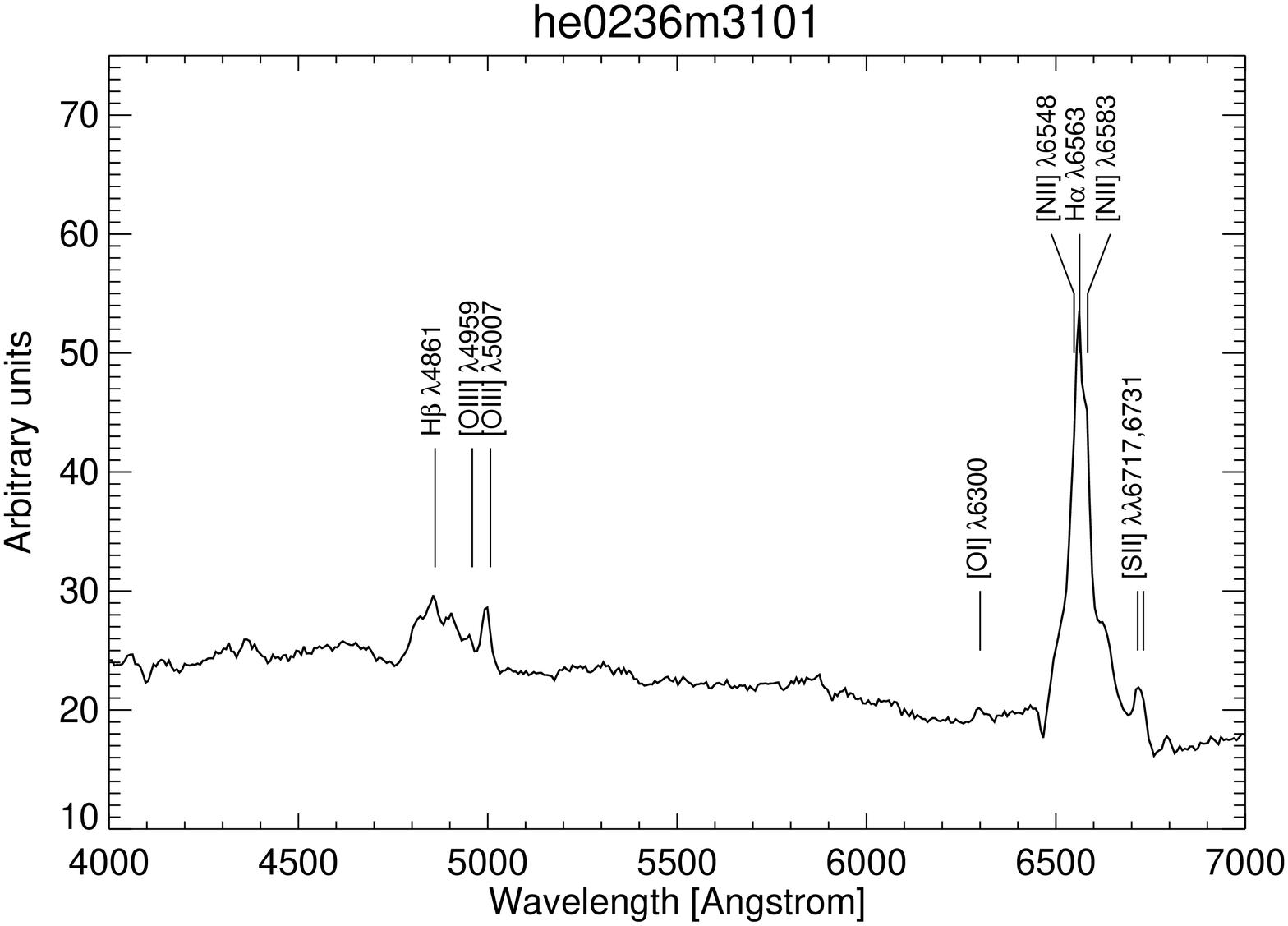}}
\subfloat[]{\label{fig:hb}\includegraphics[width=0.5\textwidth]{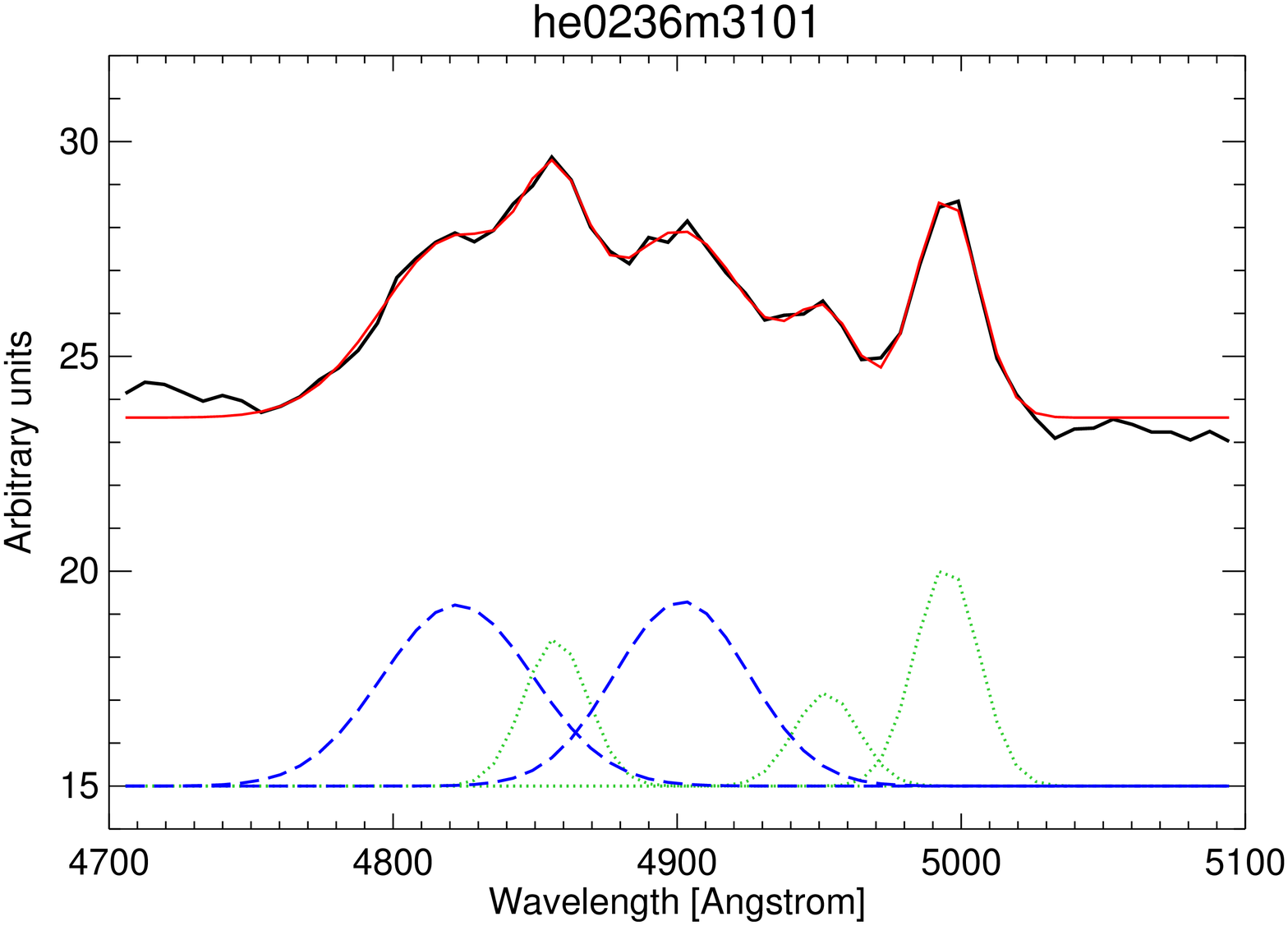}}
\caption[The galaxy HE0236-3101 and its optical spectrum and the fitting of lineH$\beta$ using the HES data.]
{\label{fig:fittinghb}(a) HE0236-3101 (ID~27) as observed by the HES. All emission lines of interest are labeled. (b) Fitting of H$\beta$ $\lambda$4861\AA{}. Note the double H$\beta$ broad components needed to fit the spectrum. The blue dashed lines represent the broad components, the green dotted lines represent the narrow components, and the solid red line represents the final fit.}\label{fig:fig4}  
\end{figure*}

The H$\beta$ and [O\,{\sc{iii}}] lines were analyzed simultaneously (see Fig.\,\ref{fig:fittinghb}). The H$\beta$ broad component was fitted with no constraints. We assumed the three narrow components  have the same width and  kinematics. Additionally, we fit the [O\,{\sc{iii}}] lines  with two Gaussians with the intensity ratio fixed to the theoretical value of 3 \citep{2007MNRAS.374.1181D}.

\begin{figure*}
\centering\subfloat[]{\label{fig:allha}\includegraphics[width=0.5\textwidth]{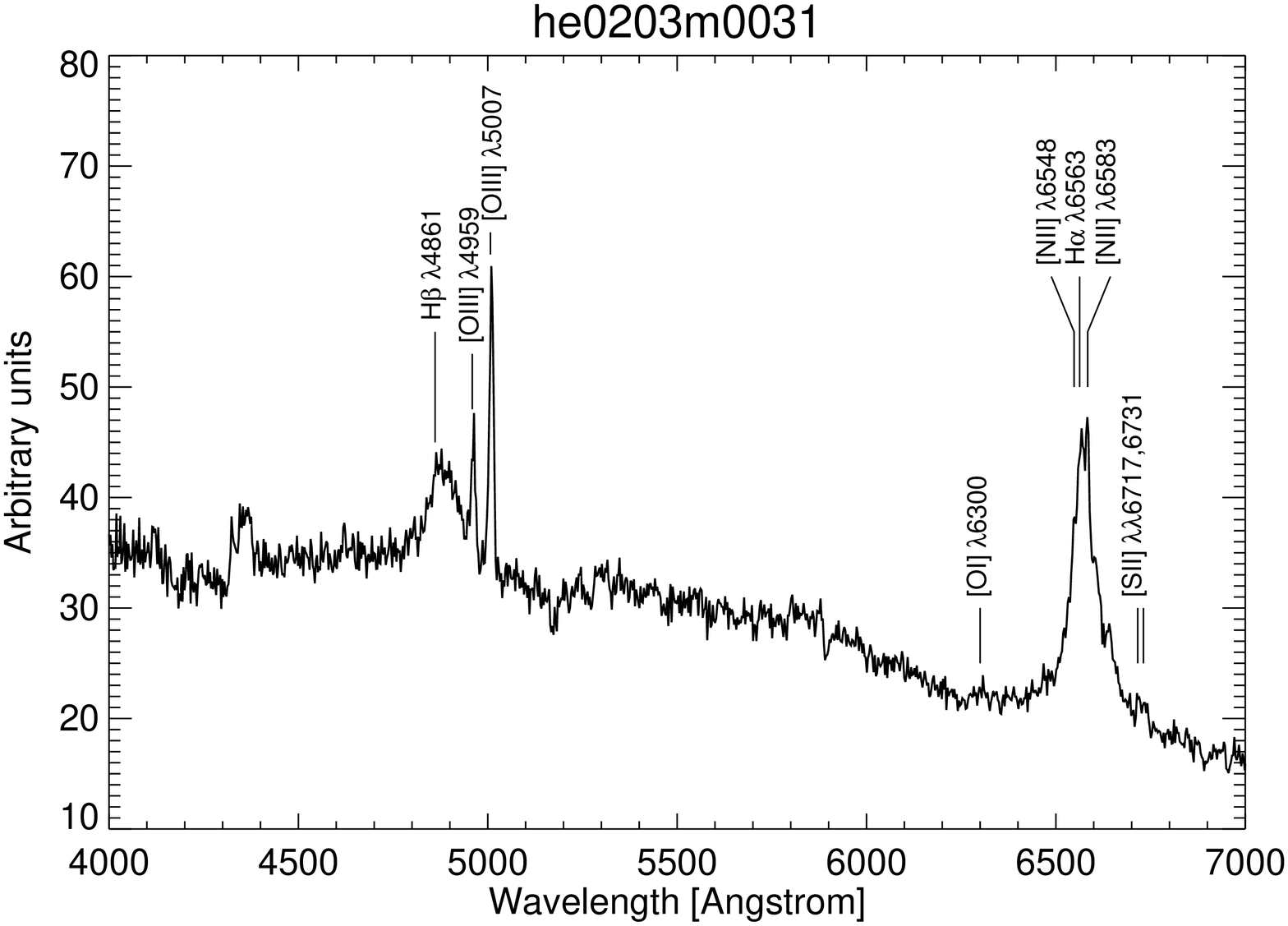}}
\subfloat[]{\label{fig:ha}\includegraphics[width=0.5\textwidth]{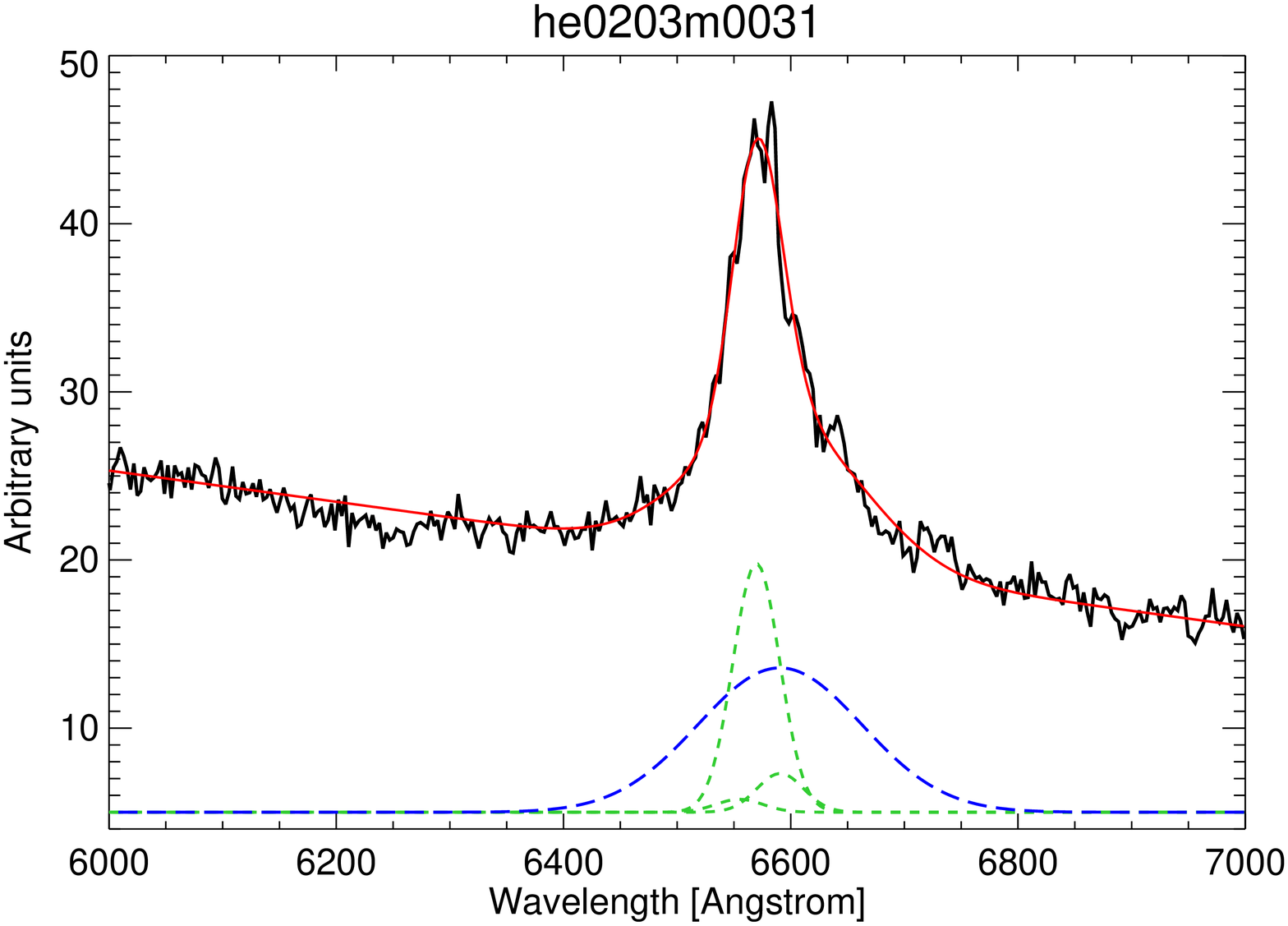}}
\caption[The galaxy HE0203-0031 and its optical spectrum and the fitting of lineH$\alpha$ using the HES data.]
{\label{fig:fittingha}(a) Same as Fig.\,{\ref{fig:fittinghb}, except for galaxy  HE~0203-0031 (ID~22) as observed by the HES}. (b) Fitting of H$\alpha$ $\lambda$6562\AA{} with one broad component for H$\alpha$ and three narrow components for the H$\alpha$- N\,[{\sc{ii}}] complex. The blue dashed line represents the broad component, the green dotted lines respresent the narrow components, and the solid red line represents the final fit.}\label{fig:fig5} 
\end{figure*}

\begin{figure*}
\centering
\subfloat[]{\label{fig:alloi}\includegraphics[width=0.5\textwidth]{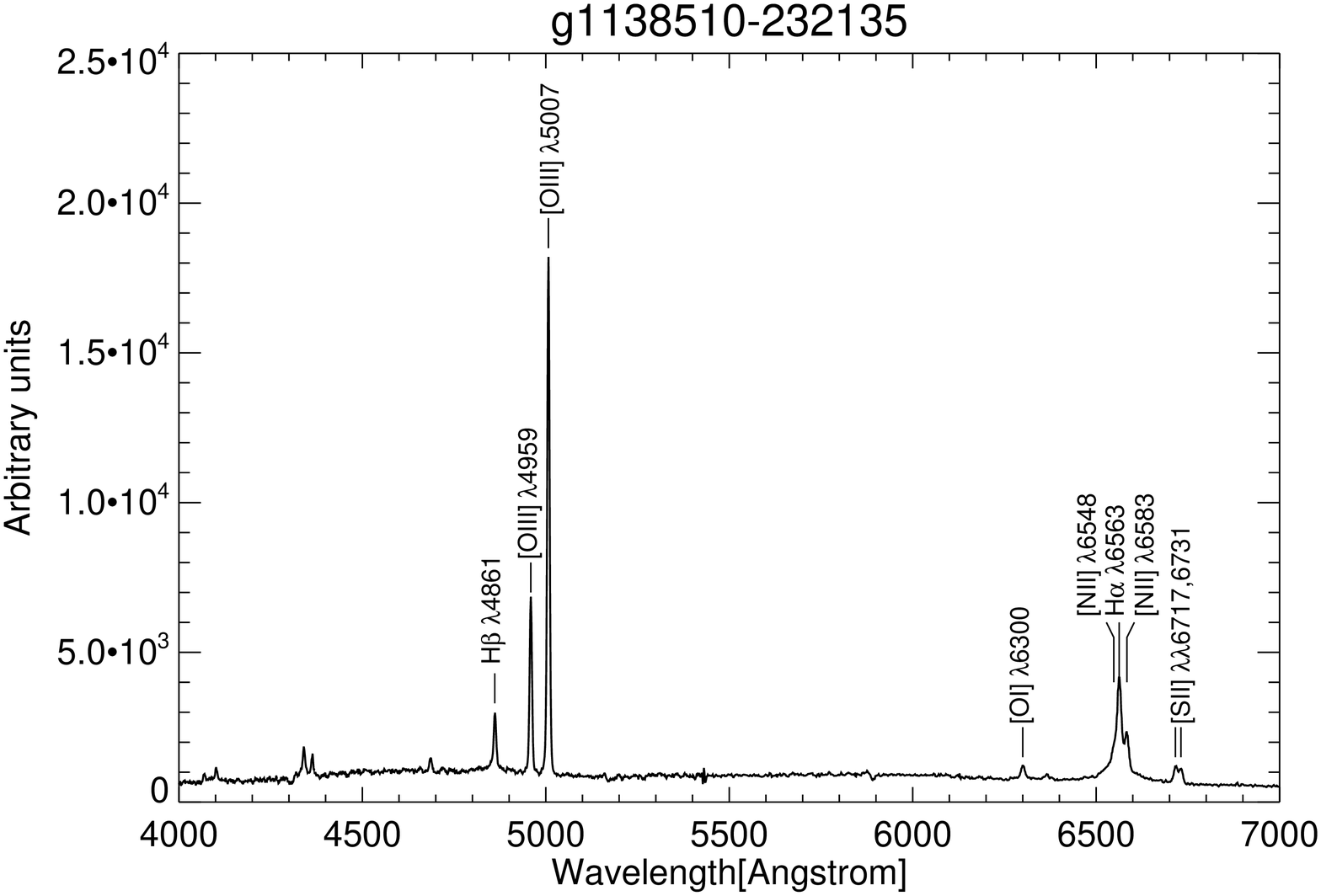}}
\subfloat[]{\label{fig:oi}\includegraphics[width=0.5\textwidth]{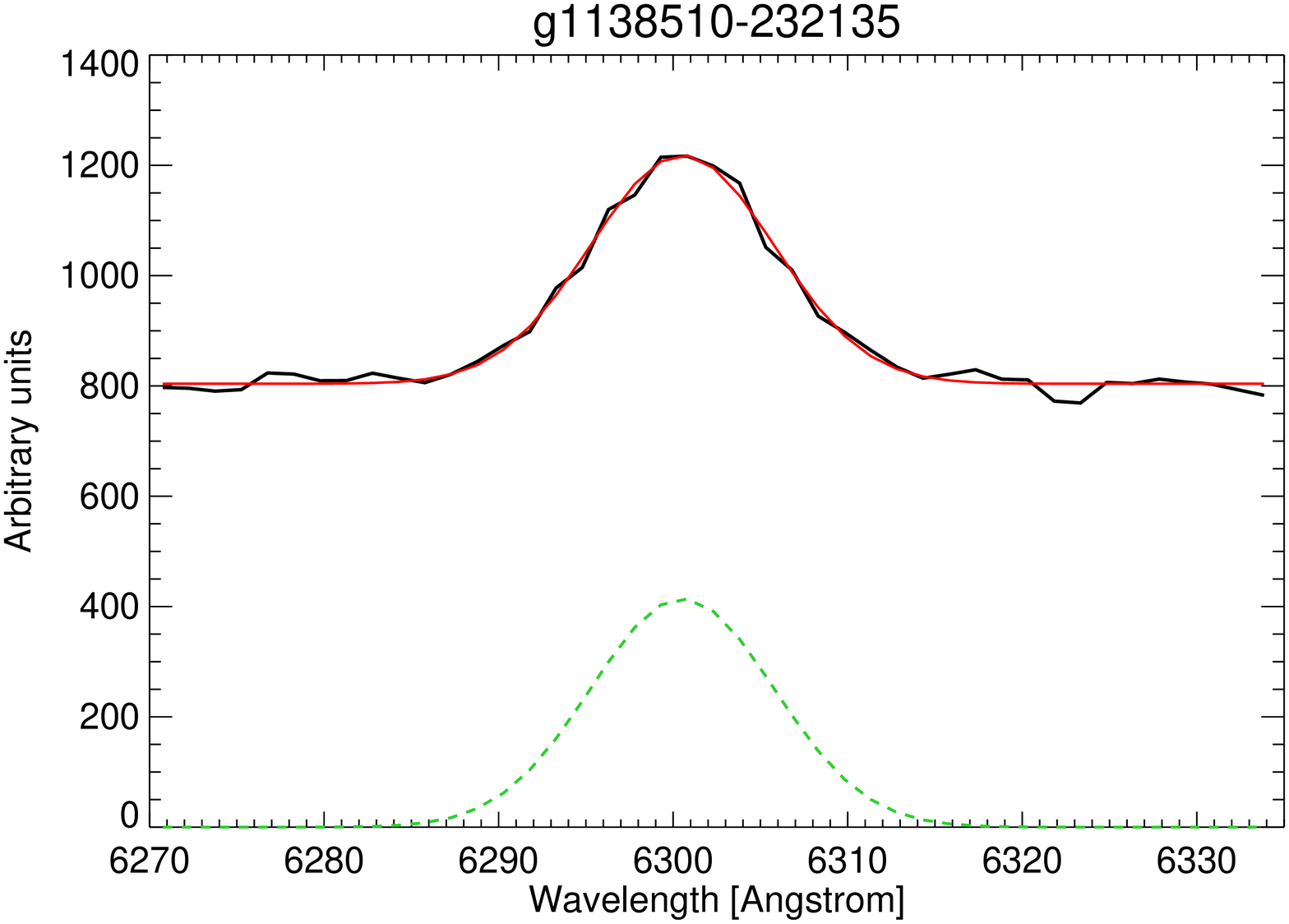}}
\caption[The galaxy g1138510-232135 and its optical spectrum and the fitting of OI $\lambda$ 6300\AA{}.] 
{\label{fig:fittingoi}(a) 6dFGS rest-frame optical spectrum of HE~1136-2304 (ID~66). The optical emission lines of interest are labeled. (b) Fitting of [O\,{\sc{i}}] $\lambda$ 6300\AA{}.}\label{fig:fig3}  
\end{figure*}

\begin{figure*}
\centering\subfloat[]{\label{fig:allsii}\includegraphics[width=0.5\textwidth]{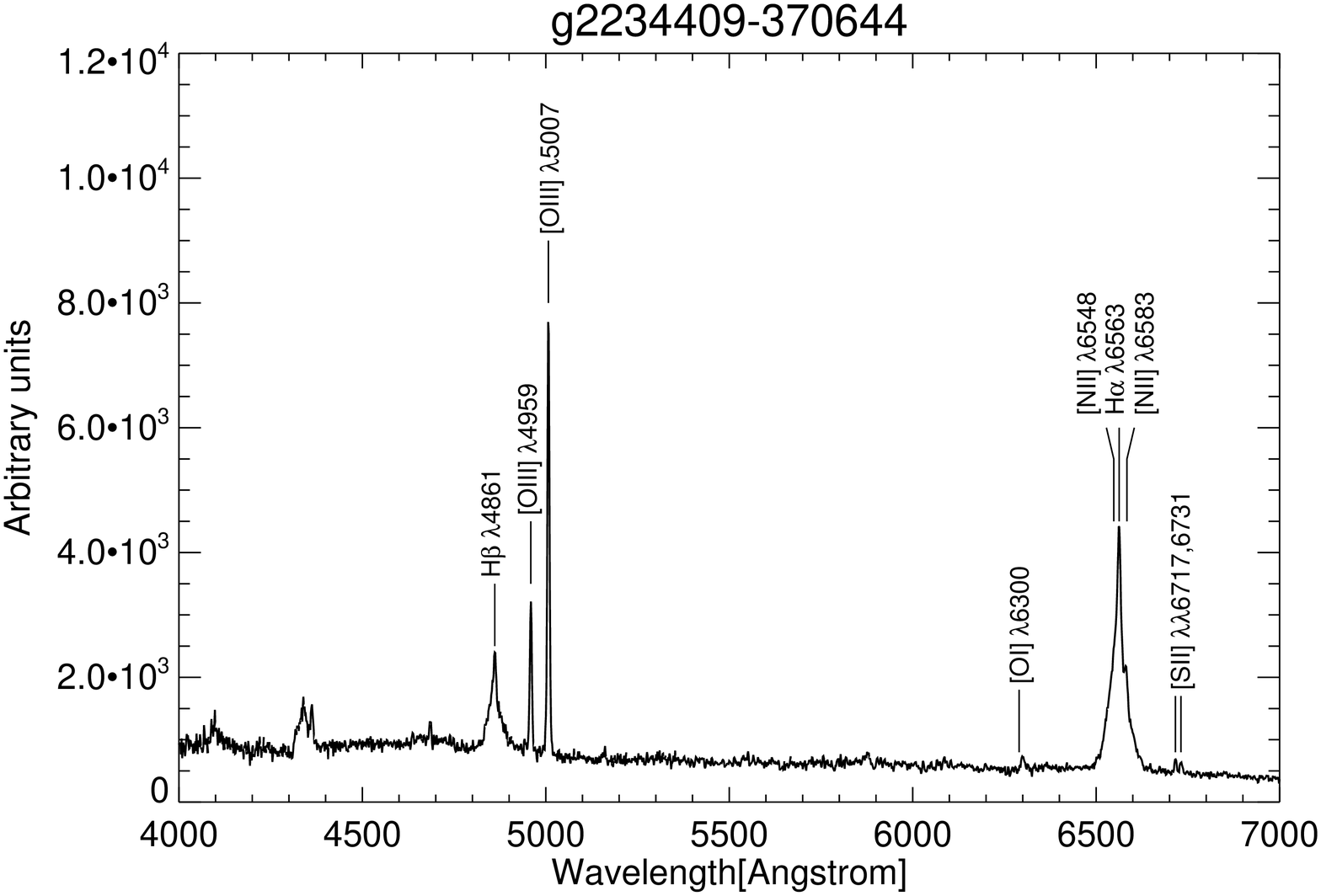}}
\subfloat[]{\label{fig:sii}\includegraphics[width=0.5\textwidth]{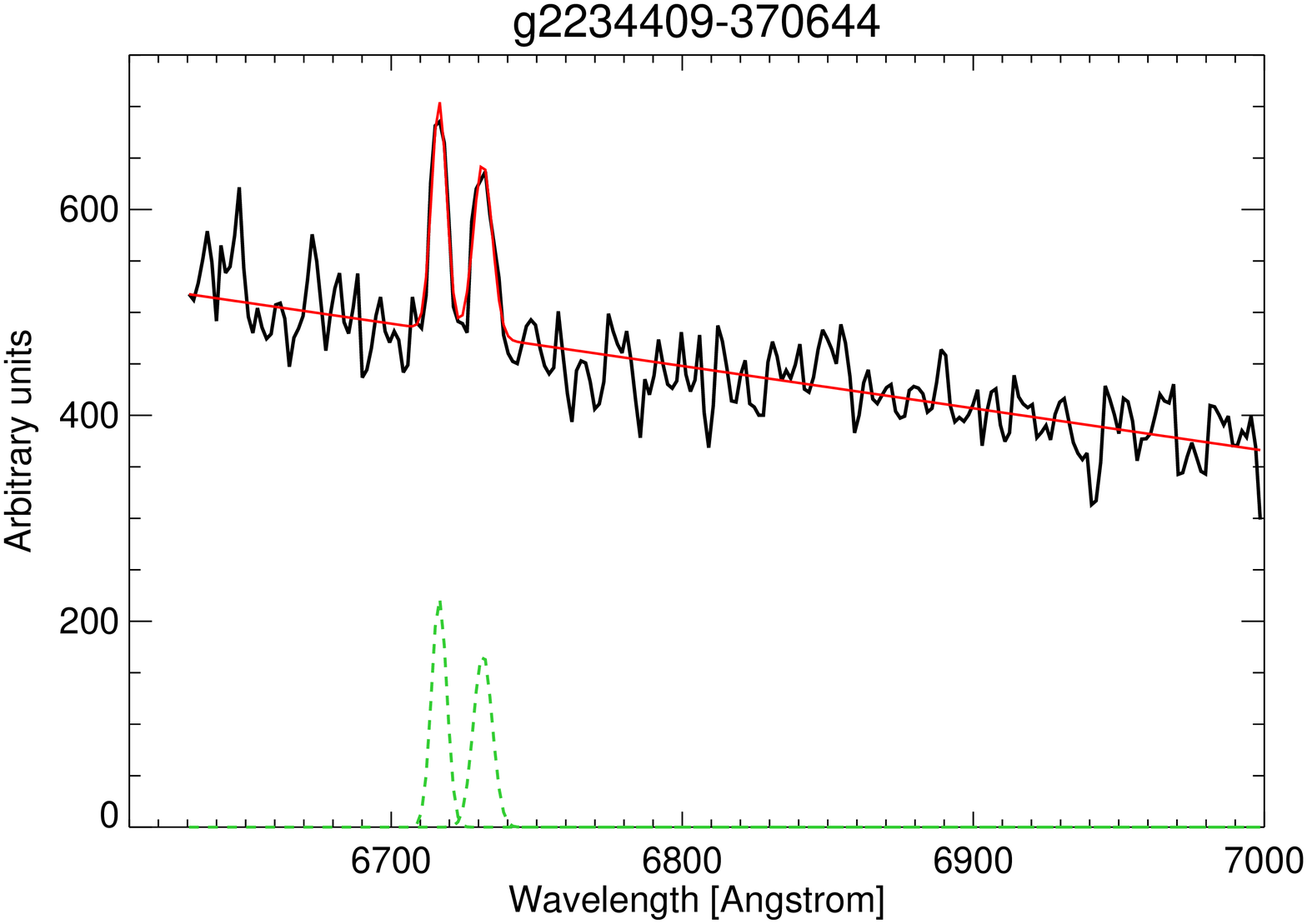}}
\caption[The entire spectrum of the galaxy g2234409-370644 and the fitting of line S{\sc{ii}} $\lambda$ 6717\AA{} and $\lambda$  6731\AA{}] 
{(a) Same as Fig.\,{\ref{fig:fig3}, except for galaxy HE~2231-3722 (ID~87)}. (b) Fitting of [S\,{\sc{ii}}] $\lambda$ 6717\AA{} and $\lambda$  6731\AA{}.}\label{fig:fittingsii}
\end{figure*}

Similarly, the H$\alpha$ and [N\,{\sc{ii}}] lines complex was simultaneously fitted (Fig. \ref{fig:fittingha}), and the H$\alpha$ broad component was left free. The narrow components were assumed to have the same kinematics and width. For the [N\,{\sc{ii}}] doublet, the intensity ratio was set to 3 as well \citep{1989agna.book.....O, 1997ApJS..112..391H}.
The [O\,{\sc{i}}] lines (Fig.\,\ref{fig:fittingoi}) were individually fitted with a single component, with no particular constrains. Finally, as the [S\,{\sc{ii}}] line ratio can be used to estimate the electron density of the emitting gas \citep[e.g.,][]{1989agna.book.....O}, we only assumed that both lines share the same kinematics (see Fig. \ref{fig:fittingsii}).

With the intention of having a more robust result, we tried to impose the same FWHM to all narrow emission lines. This approach did not work, because in many cases, especially for the HES data, strong blending was affecting the H$\alpha$+[N\,{\sc{ii}}] complex. This made it very difficult for the fit to converge, especially considering that in most cases an extra broad component was needed to fit H$\alpha$. Because of this and the inherent difficulties of the performed fits in the H$\beta$ region due to the presence of Fe\,{\sc{ii}} lines and uncertainties in the result of the spectral slope, we chose not to impose conditions between different spectral regimes. Instead, we  demonstrated that the results were consistent for both narrow and broad components after the fits were finished (see Section 5).

\section{General fitting results}\label{fitres} 
We have analyzed the optical spectra of 71 observed with the
HES (long-slit spectroscopy) and 58 sources obtained with the 6dFGS (fiber spectroscopy). 
Tables \ref{tab:6df} and \ref{tab:hes} give information on the emission line flux ratios and line widths for 6dFGS and HES data, respectively. Because of low SNR in the HES spectra, in most cases the [S\,{\sc{ii}}] and [O\,{\sc{i}}] spectral fits did not converge. As a result, for our analysis of the HES data set we only used the Balmer, [N\,{\sc{ii}}], and [O\,{\sc{iii}}] emission lines.

\subsection{Width of  emission lines }\label{emibroad}

\begin{figure*}
\centering \subfloat[]{\label{fig:broad_hes}\includegraphics[width=0.45\textwidth]{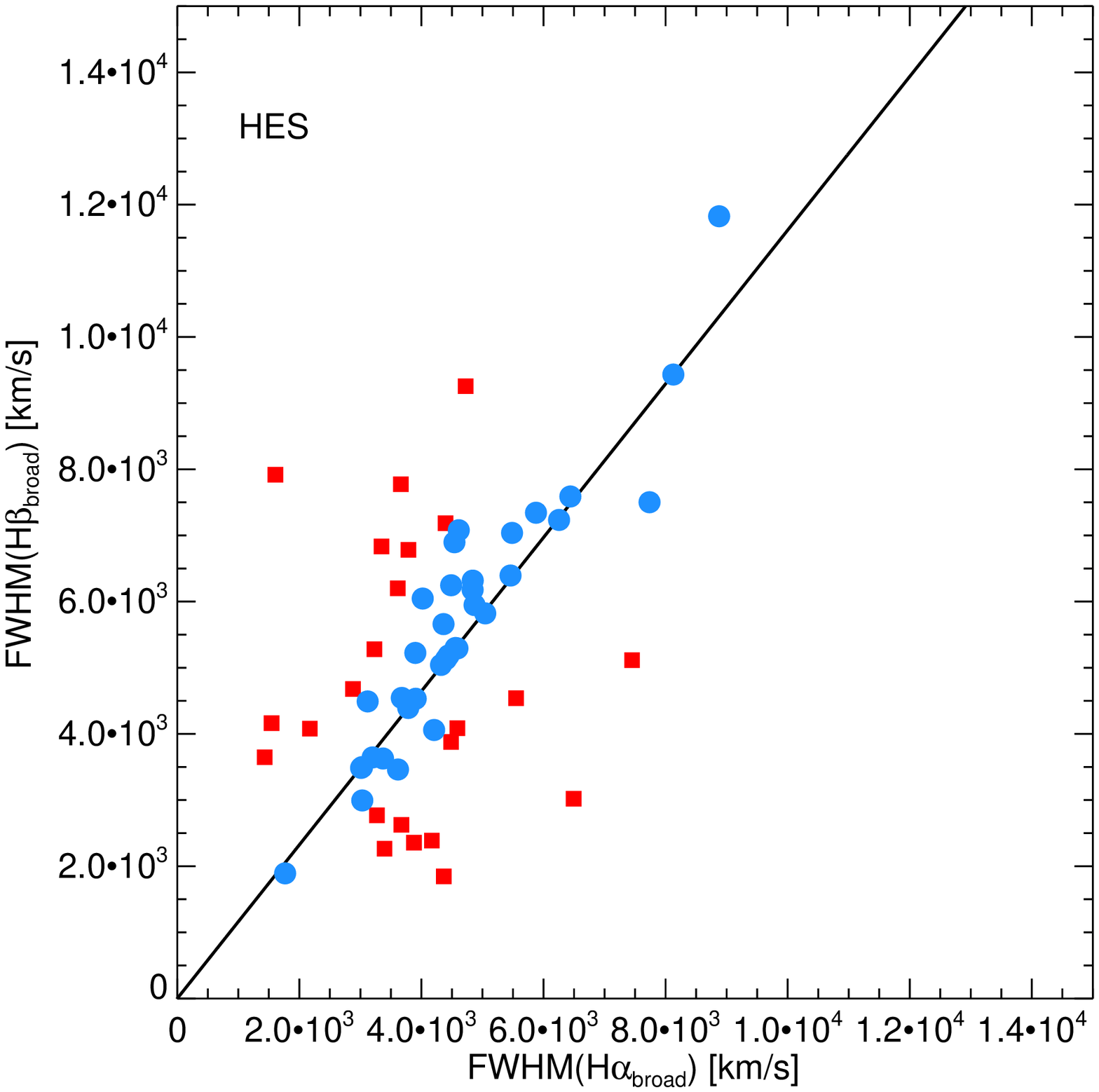}}
\subfloat[]{\label{fig:broad_6df}\includegraphics[width=0.45\textwidth]{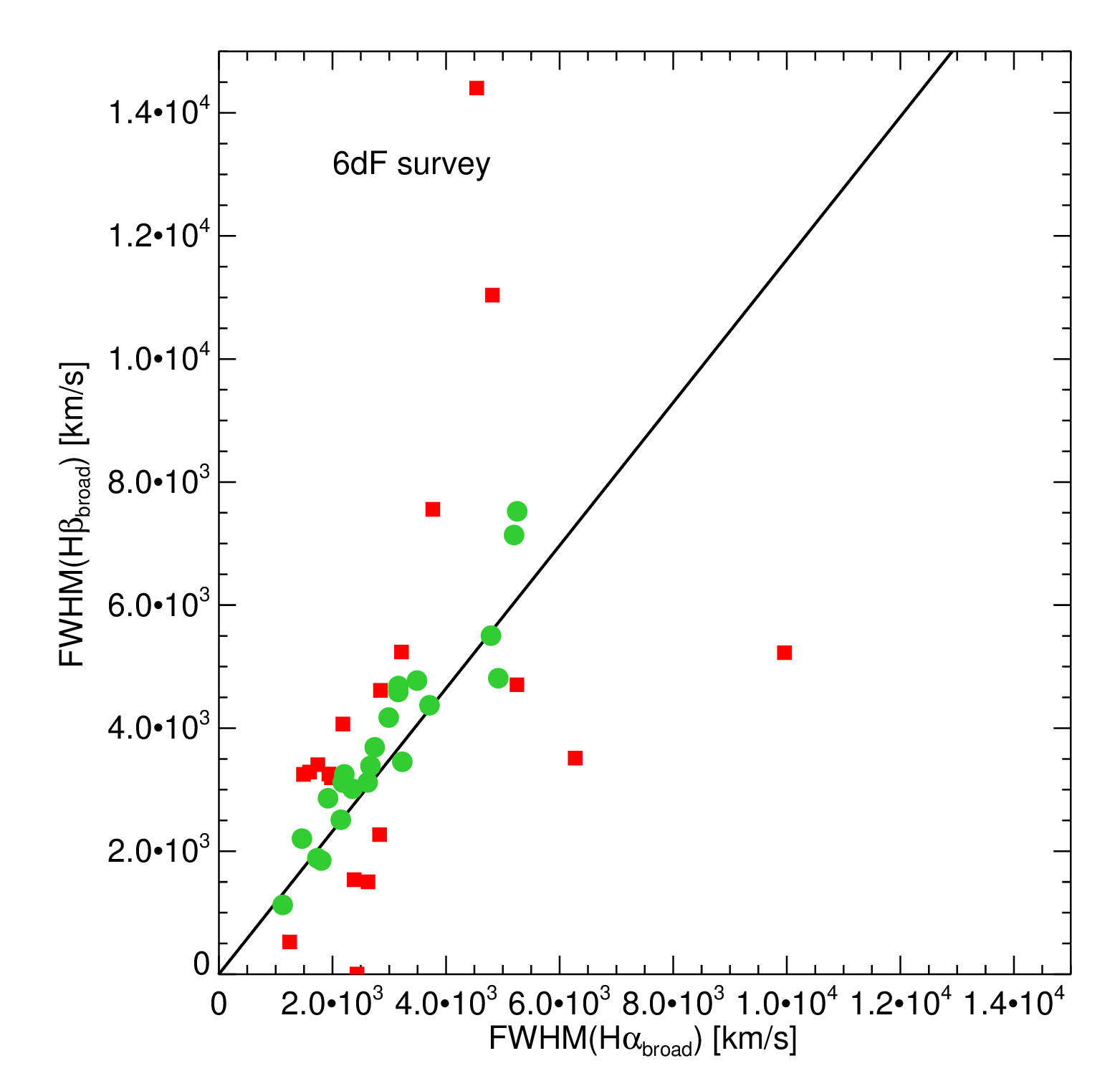}}
\caption[FHWM broad components for H$\alpha$ and H$\beta$]
{\label{fig:broad_both} FWHM of broad components fitted to the hydrogen recombination lines. The black line represents the expected relation (FWHM(H$\alpha$) = 0.861FWHM(H$\beta$)) as given by \citet{Zhou2006}. Squares represent galaxies showing departures from this behavior of more than 25\%. Circles represent galaxies that comply with this relation within a 25\%.}(a) Results for the HES. (b) Results for the 6df survey.
\end{figure*}

\begin{figure*}
        \centering\subfloat[]{\label{fig:narrow_hes}\includegraphics[width=0.45\textwidth]{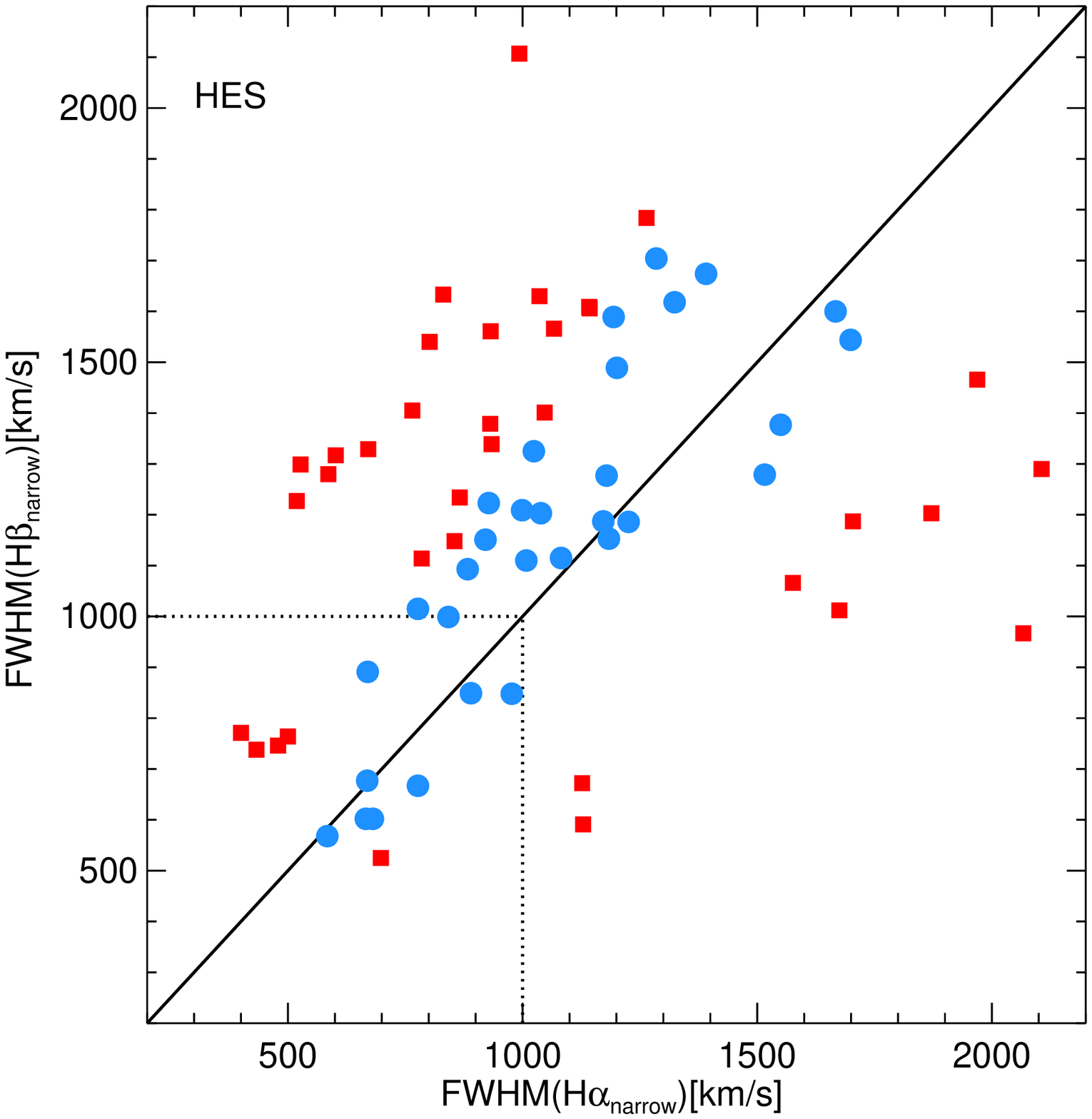}}
        \subfloat[]{\label{fig:narrow_6df}\includegraphics[width=0.45\textwidth]{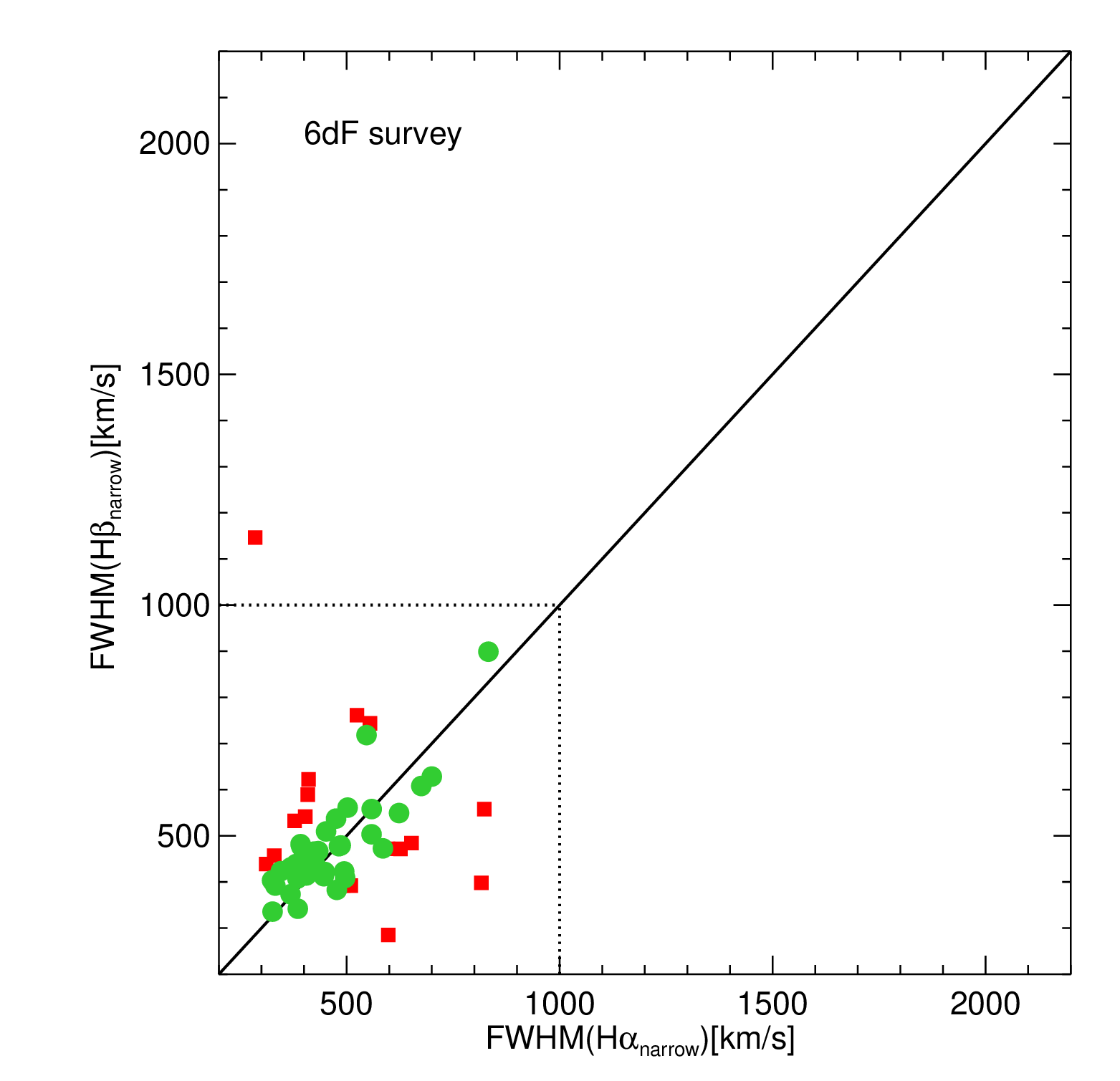}}
        \caption[FHWM narrow components for H$\alpha$ and H$\beta$.]
{       \label{fig:narrow_both} FWHM of narrow components fitted to the hydrogen recombination lines. The black line represents the ideal behavior (FWHM(H$\alpha$) = 1.0 FWHM(H$\beta$)). Squares represent galaxies showing departures from this behavior of more than 25\%. Circles represent galaxies that comply with this relation within a 25\%.(a) Results for the HES. (b) Results for the 6df survey.} 
\end{figure*}

The 6dFGS line widths have been deconvolved according to the spectrometer resolution. For the HES sample we are giving observed widths; this is because a variety of different resolutions were used in the observations. The HES widths should hence be considered an upper limit. In this particular case this is not critical, as the integrated flux would not change and we are using the width information only to characterize and give a quality flag to the spectral fits.\\
Figure\,\ref{fig:broad_both} compares the FWHM values of the fitted broad hydrogen recombination lines components, for both samples. All sources have FWHM larger than 1000 km s$^{-1}$, but the values are very scattered and do not follow any trend. Based on a study of more than 1000 local (z $\lesssim$ 0.4) NLSy1, \citet{Zhou2006}  found that the widths of the H$\alpha$ and H$\beta$ broad components do follow a linear correlation. They find the best fit to be: $FWHM(H\alpha) = 0.861 \times FWHM(H\beta)$. This relation has been positively tested for normal Seyfert 1 galaxies (M. Valencia-Schneider, priv. comm.), and reflects the fact that the lines are coming from different layers of the BLR.
Forty-three\% and 51\% of the 6dFGS and HES galaxies follow the expected behavior within a 25\% dispersion (see blue  and green circles in Figure\,\ref{fig:broad_both}). Although the HES spectrometer resolution is different for the H$\alpha$ and H$\beta$ spectral regimes, we assume that the considered 25\% uncertainty takes that into consideration. Both H$\beta$ and H$\alpha$ narrow components should have similar line widths, as their emission comes from the same region and trace identical mechanisms. In this case, 68\% and 44\% of the 6dfGS and HES, repectively, exhibit the same FWHM for both narrow lines within a 25\% deviation (see blue and green circles in Fig.\,\ref{fig:narrow_both}). The mean FWHM and standard deviations of both full samples and selected galaxies are given in Table\,\ref{tab:averages}.
As the spectral resolution in this investigation is limited, the diagnostic emission line
ratios may still contain hidden contributions of broad lines in H$\alpha$ and
H$\beta$. However, as these lines are broad, and probably of low intensity, they only
have a minor influence on the diagnostic ratios and on the conclusions we base them on.
\\
These departures in line widths measurements,  between the lines of the same survey and  average survey values, are not unexpected. The differences between survey results, especially in the H$\alpha$ narrow components, can be mostly accounted for by the better spectral resolution of the 6dfGS. Also, strong line blending, uncertainties in the result of the slope of the spectra, or contamination from iron lines in the case of H$\beta$, can be responsible for  departures from the expected correlations. This has an impact on the measured fluxes, and may have an impact on the galaxies' classification. For this reason, from now on we separate the two galaxy groups (following or not the expected correlations).

\begin{table}
\begin{center}
\caption{Hydrogen recombination lines average FWHM and standard deviation for all studied galaxies, and for the galaxies selected following the critieria defined in Section \ref{emibroad}.}\label{tab:averages}
    \begin{tabular}{ | l | c | c | c | c |}
    
    \hline
    Line  & 6df [km/s] & HES [km/s] & 6df [km/s] & HES [km/s]  \\ 
          & All Gxs. & All & Sel. & Sel. \\ \hline
    H$\beta_{Broad}$   & 4120$\pm$2545 & 5179$\pm$2012 & 3754$\pm$1563 & 5610$\pm$1917 \\ \hline
    H$\beta_{Narrow}$  &  501$\pm$147  & 1184$\pm$349  &  479$\pm$110  & 1156$\pm$331  \\ \hline
    H$\alpha_{Broad}$  & 3783$\pm$4431 & 4301$\pm$1806 & 2926$\pm$1180 & 4623$\pm$1490 \\ \hline
    H$\alpha_{Narrow}$ &  487$\pm$140  & 1093$\pm$527  &  469$\pm$114  & 1065$\pm$300  \\ 
    \hline
    \end{tabular}
\end{center}
\end{table}

\subsection{Narrow line Seyfert 1 subsample}\label{nls1}
Classically, narrow line Seyfert galaxies (NLS1s) are defined by the
width of their optical Balmer emission lines, such as H$\beta$, in combination
with the relative weakness of the [O\,{\sc{iii}}] $\lambda 5007\AA$ emission
line, i.e. FWHM of the broad H$\beta$ component less than 2000 km/s and
[O\,{\sc{iii}}] / H$\beta_{\rm total}< 3$
\citep[e.g.,][]{1985ApJ...297..166O,1989ApJ...342..224G, 2008RMxAC..32...86K}.
There is some controversy as to whether NLS1 are a special class of AGN or an extension of classical Type 1 objects.  \citet{Valencia2012} found that the observed parameters in Type 1 AGN are continuous, with no particular difference between sources with FWHM(H$\beta_{broad}$) above or below 2000 km/s. This fact is inconsistent with the existence of two populations. We, however, chose to find out the relevance of the NLS1 galaxies populations, to compare with the results in the literature.

In our sample, and based on the 6dFGS data, we detect six galaxies that fulfill the classic NLS1 requirements (IDs 5, 8, 29, 54, 77, 91, see also Table\,\ref{tab:6df}). At this point, we do not consider the often associated Fe\,{\sc{ii}}
emission \citep[cf. ][]{2001A&A...372..730V}.
Two out of six NLS1 galaxies are consistent with previous references as NLS1 (IDs 5,77 Gruppe et al.2003 and Dietrich et al. 2005, respectvely) in the literature, while one of them is classified as Sy1.9 \citep[ID$\sim$29 in Tab.
\ref{tab:6df};][]{veron}. Three galaxies of the NLS1 group have not been
classified as such before (IDs 8,54,91). 

\subsection{Population A vs. Population B}
AGNs are separated into  radio-loud and radio-quiet objects, but to our knowledge there has not been any deep analysis or dedicated observations to verify the radio loudness of the sample under study.
In the context of optical emission line properties, Seyfert1, NLS1, Seyfert2, QSOs, and broad absorption line QSO and LINERs are all radio-quiet galaxies. The radio-loud galaxies can  be broadly divided into low-excitation and high-excitation classes \citep{Hine1979,Laing1994}. Low-excitation objects do not show both broad and strong-narrow emission lines. They can exhibit weak narrow emission lines that may be excited by a different mechanism  \citep{Baum1995}; their optical and X-ray nuclear emission are consistent with being generated in a jet \citep{Hardcastle2006}. Comparatively, the optical emission line spectra of high-excitation objects (narrow-line radio galaxies) is similar to those of Seyfert 2 galaxies. 
The small class of broad-line radio galaxies, which show relatively strong nuclear optical continuum emission (Grandi \& Osterbrock 1978) and likely compress some low-luminosity, radio-loud quasars.
Based on this, and with no radio information, the present sample should be comprised of radio-quiet galaxies, radio-loud high excitation and broad line radio galaxies.
\citet{Sulentic2000, Sulentic2000b}  identified two radio-quiet AGN populations. Population A is an almost purely radio-quiet, with FWHM$\leq$4000 km s$^{-1}$, generally strong Fe{\sc{ii}} emission and a soft X-ray excess. Comparatively, Population B has FWHM$\geq$4000 km s$^{-1}$ and optical properties largely indistinguishable from flat spectrum radio-loud sources, including usually weak Fe{\sc{ii}} emission. A possible interpretation sees population A as lower BH mass/high accretion rate sources and population B/radio-loud sources as the opposite.
Because of the better spectral resolution and accuracy on the spectral fits, we  used the 6dFGS data of our sample to study the prevalence of Populations A and B. Twenty-four of the galaxies with H$\beta$ broad component have FWHM$\lesssim$4000 km s$^{-1}$, consistent with Population A, 
33\% of them have strong Fe{\sc{ ii}} emission. In addition 12\% have weak Fe{\sc{ ii}}. As for Population B sources, 17 galaxies have FWHM$\gtrsim$4000 km s$^{-1}$, with 30\% having weak and 18\% strong Fe{\sc{ ii}} emission. Fifty\% of our sources with H$\beta$ broad emission fit well into the categories designated by \citet{Sulentic2000, Sulentic2000b}. We hypothezise that the remaining 50\% are low-luminosity, radio-loud quasars.


\subsection{Galaxies with double broad components}
Two out of the 87 galaxies with HES data, HE0236-3101 (ID 27, see Fig.\,\ref{fig:hb}) and HE0236-5224 (ID 28),
 give a better H$\beta$ fit result when using a double broad component. Already \citet{1991ApJ...382L..63C} found the double broad component in HE~0236-3101, whereas the case of HE0236-5224 has never been described before. 
 No 6dFGS data was available for these galaxies, so further emission lines cross checking was not possible.\\

\subsection{Galaxies with no broad components}
A few galaxies do not present one or both of the expected broad emission lines associated with the hydrogen recombination lines. When both broad lines are not detected, we classify those sources as no broad line emitters (N.B.E in Tables\,\ref{tab:6df} and \ref{tab:hes}). We find two 6dFGS (sources ID 13 and 51) and one HES sources (ID 51) classified as N.B.E., which makes a total number of two sources in the entire sample. This means that the sources are not Type 1 Seyferts, or that the broad components are too weak and buried in the noise.
Tables\,\ref{tab:6df} and \ref{tab:hes} give details on other sources in which one broad component, either H$\alpha$ or H$\beta$, is not detected. In those cases the nondetections are likely to be lines with low SNR, or lines subject to strong line blending (in particualr, for the H$\alpha$ lines in HES sources) that makes the fit more uncertain.

\subsection{The NLR electron density}
\begin{figure}[h]
\begin{center}
\includegraphics[width=\columnwidth]{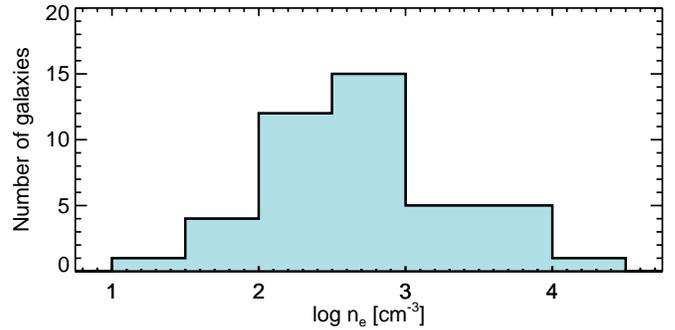}
\end{center}
\caption{Distribution of the electron density as estimated from the [S\,{\sc{ii}}]$\lambda$6716/[S\,{\sc{ii}}]$\lambda$6731 using the function given by \citet{Osterbrock2006}.}
\label{fig:electron_density_hist}
\end{figure}

We derived the electron density of 43 sources of the 6dFGS, using the [S\,{\sc{ii}}]$\lambda$6716/[S\,{\sc{ii}}]$\lambda$6731 lines ratio as a function of electron density at 10$^{4}$K (see \citet{Osterbrock2006}). The results are clearly skewed, with values clustered toward lower densities, between 100 and 1000 N$_{e}$/cm$^{3}$ (see Figure \ref{fig:electron_density_hist}). Densities of the order of $\log \left(\frac{N_{e}}{{\rm cm}^{-3}}\right) \sim 2.2$ are typical for the narrow-line regions of broad line AGN \citep[e.g.,][]{2007ApJ...670...60X}. 


\section{Optical emission lines diagnostic diagrams for the nearby LLQSOs sample}\label{class}
\begin{figure}[h]
  \centering
  \includegraphics[width=7cm]{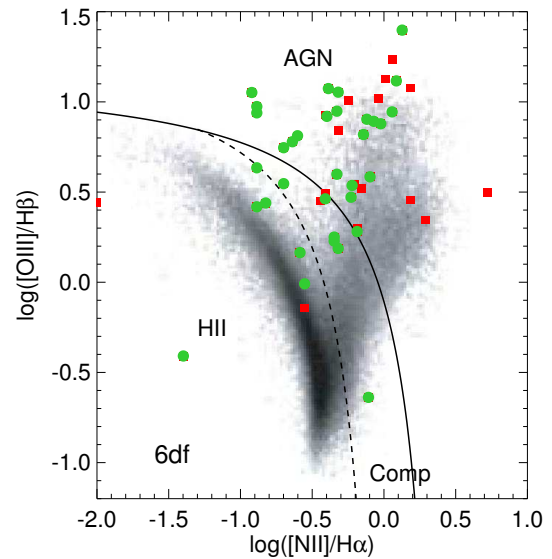}
      \caption{[N\,{\sc{ii}}]/H$\alpha$ versus [O\,{\sc{iii}}]/H$\beta$
      diagnostic diagram for the 6dFGS data. The solid line
      is the theoretical maximum starburst limit (Kewley et al. 2001). 
      The sources that are located
      above the line are classified as AGN. Galaxies that lie below the star
      formation demarcation dashed line (Kauffmann et al. 2003a) are
      classified as H\,{\sc{ii}} region-like galaxies. The galaxies that lie
      between the two demarcation lines are on the AGN-H\,{\sc{ii}} mixing
      sequence and are called composite galaxies. The dark cloud represents
      SDSS galaxies from a previous study by \cite{kewley2006}. Green circles represent 
      galaxies that follow the narrow lines width criteria (see Sect.\,\ref{emibroad}), and red squares represent those that do not.}
         \label{fig:6dfnii}
   \end{figure}

\begin{figure}[h]
  \centering
  \includegraphics[width=7cm]{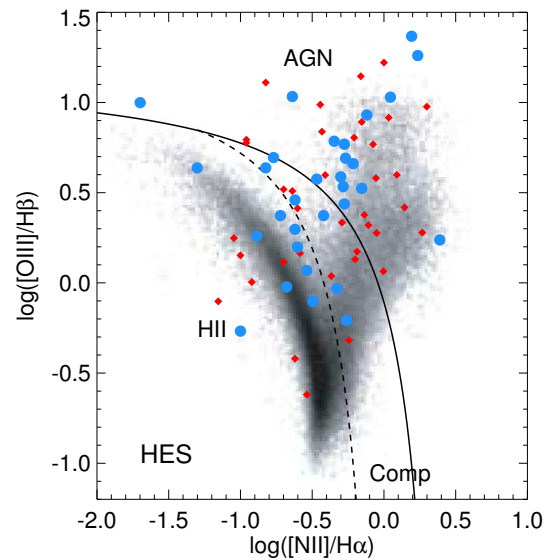}
      \caption{Same as Fig.\,\ref{fig:6dfnii}, except with the HES data. Blue circles represent 
      galaxies that follow the narrow lines width criteria (see Sect.\,\ref{emibroad}), and red squares represent those that do not.}
         \label{fig:hesnii}
   \end{figure}

\begin{figure}[h]
  \centering
  \includegraphics[width=7cm]{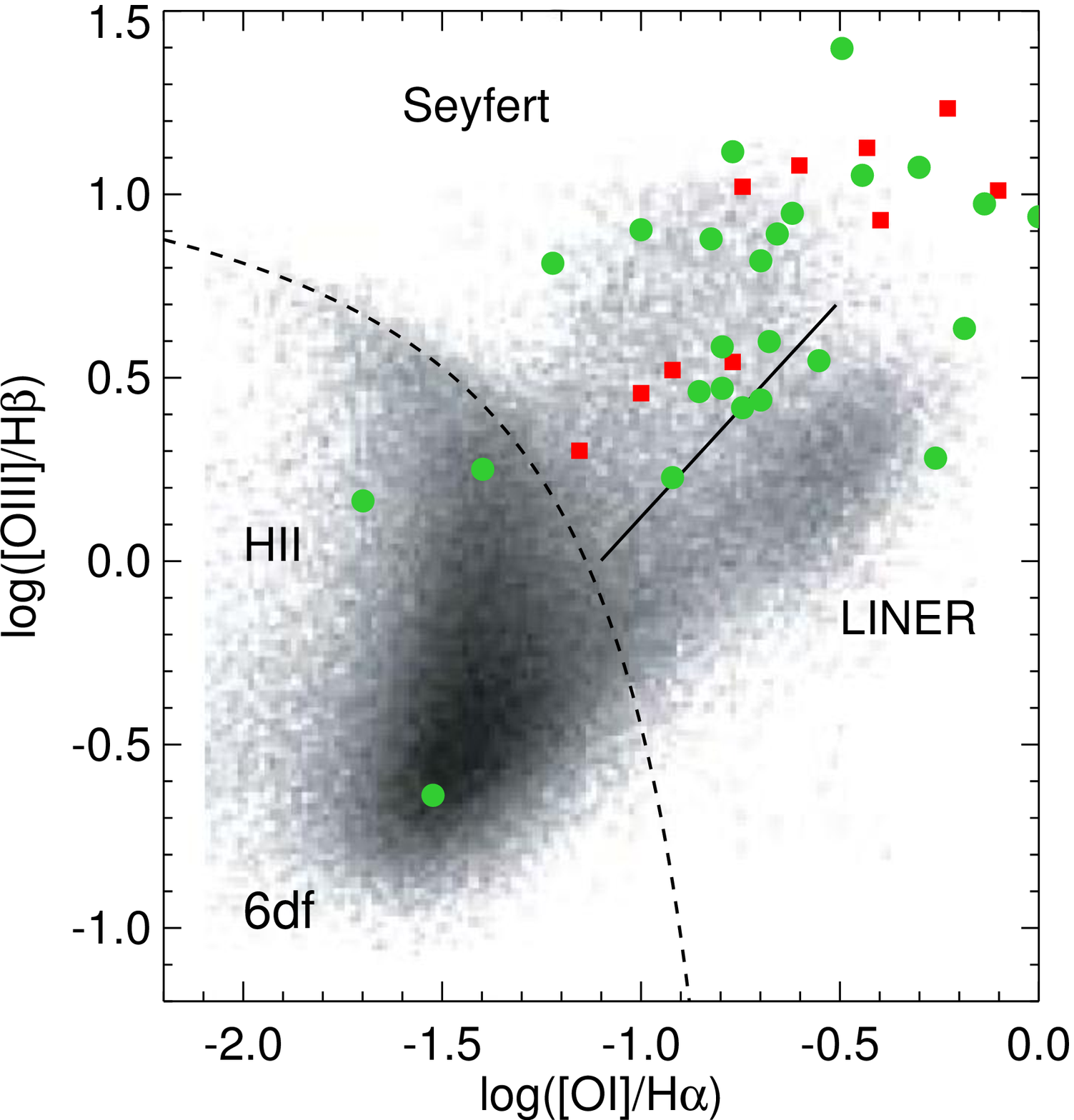}
      \caption{[O\,{\sc{i}}]/H$\alpha$ versus [O\,{\sc{iii}}]/H$\beta$
      diagnostic diagram for the 6dFGS data. The demarcation lines are from
      \cite{kewley2006}. The solid line represents the division between
      Seyferts and LINERs.  The dark cloud represents SDSS galaxies from a
      previous study by \cite{kewley2006}. Green circles represent 
      galaxies that follow the narrow lines width criteria (see Sect.\,\ref{emibroad}), and red squares represent those that do not.}
         \label{fig:6dfoi}
   \end{figure}

\begin{figure}[h]
  \centering
  \includegraphics[width=7cm]{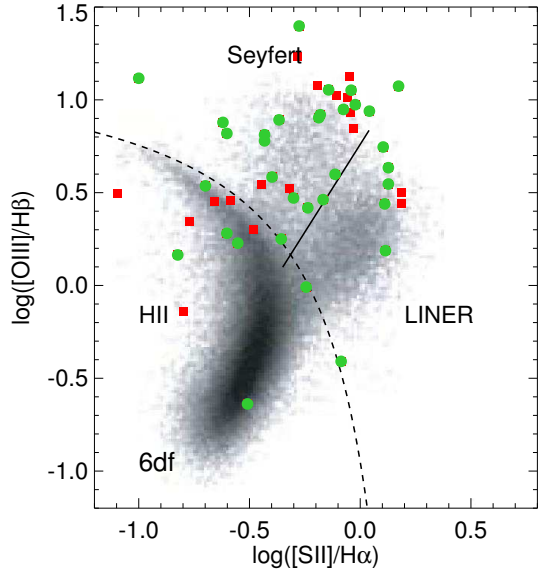}
      \caption{[S\,{\sc{ii}}]/H$\alpha$ versus [O\,{\sc{iii}}]/H$\beta$
      diagnostic diagram for the 6dFGS data.  The demarcation lines are from
      \cite{kewley2006}. The solid line represents the division between
      Seyferts and LINERs.  The dark cloud represents SDSS galaxies from a
      previous study by \cite{kewley2006}. Green circles represent 
      galaxies that follow the narrow lines width criteria (see Sect.\,\ref{emibroad}), and red squares represent those that do not.}
         \label{fig:6dfsii}
   \end{figure}

The quality of the 6dFGS spectra is sufficiently good to
clearly detect all optical emission lines involved in the three classic BPT \citep{baldwin1981}{} diagnostic diagrams. 
For the HES spectra, the detection of the [O\,{\sc{i}}] and [S\,{\sc{ii}}] lines is
quite challenging and the fitting results uncertain, hence we do not present them here. 
Figs. \ref{fig:6dfnii} and \ref{fig:hesnii}
illustrate the [O\,{\sc{iii}}]/H$\beta$ versus [N\,{\sc{ii}}]/H$\alpha$
diagnostic diagram for our nearby sample based on the 6dFGS and HES.
Figs.\,\ref{fig:6dfoi} and \ref{fig:6dfsii} show the [S\,{\sc{ii}}]/H$\alpha$ and [O\,{\sc{i}}]/H$\alpha$ versus [N\,{\sc{ii}}]/H$\alpha$ diagramas for the 6dfGS sources. Results are compared to the distribution of the SDSS galaxies used by \cite{kewley2006}. 
The latter sample selects nearby galaxies within the redshift range $0.04 < z < 0.1$, to ensure the results are not dominated by aperture effects from the SDSS fiber spectra (3~$\arcsec$ diameter). 

\begin{table}[h]
\begin{center}
\caption{Percentages of LLQSOs classified using the BPT optical diagnostic diagrams.  Only sources following the narrow hydrogen recombination lines criteria described in Sec.\,\ref{emibroad} and represented in Figs.\,\ref{fig:6dfnii} to \ref{fig:6dfsii} are used here.}\label{tab:perc_classif}
    \begin{tabular}{ | l | c | c | c | c |}    
    \hline
    Diagram  & H\,{\sc{ii}} & AGN/Sey & Comp & LINER  \\ 
    \hline
    $[$N\,{\sc{ii}}$]$/H$\alpha_{6dfGS}$  & 18\% & 64\% & 18\% & --- \\ 
    \hline
    $[$N\,{\sc{ii}}$]$/H$\alpha_{HES}$    & 29\% & 55\% & 16\% & --- \\
    \hline
    $[$S\,{\sc{ii}}$]$/H$\alpha_{6dfGS}$  & 12\% & 68\% & ---  & 20\% \\
    \hline
    $[$O\,{\sc{i}}$]$/H$\alpha_{6dfGS}$  & 19\% & 62\% & ---  & 19\% \\
    \hline
    \end{tabular}
\end{center}
\end{table}

Consistent with their classification, most sources of the LLQSOs sample populate the higher excitation region of the diagram. Very few galaxies lie in the densely populated star-forming branches of the SDSS sample with log([O\,{\sc{iii}}]/H$\beta$) $<$ 0.0

The HES and 6dfGS results from the [N\,{\sc{ii}}]-based diagram compare well, especially considering the differences in spectral resolution and aperture used in the observations (slit widths of between $1\farcs 5$
and $2\farcs 5$ for the HES and a circular aperture of 6$\farcs$ in the 6dfGS).
The other two emission-line diagnostic diagrams are not equally sensitive to the
physical processes that are taking place. For instance, the
[N\,{\sc{ii}}]/H$\alpha$ and [S\,{\sc{ii}}]/H$\alpha$ diagrams are more
sensitive to star formation with respect to the [O\,{\sc{i}}]/H$\alpha$ diagram,
which has proven to be more sensitive to shocks
\citep[cf.,][]{2008ApJS..178...20A}. In spite of this, the percentage of galaxies classified as AGN/Seyfert remains quite constant:
in all three classification schemes, between 50\% and 68\% of the sources are classified as AGN/Seyfert (see Table\,\ref{tab:perc_classif}). The HES results show the lower AGN activity, but there the quality of the spectra may be biasing the results. Depending on the diagram used,  10 to 35\% of the LLQSOs are classified as H{\sc{ii}}. This is consistent with AGNs showing star-forming activity potentially due to the emission from the extended AGN host galaxy that falls into the aperture used. As discussed by \citet{2014MNRAS.441.2296M}, in a study using  simulations as well, the starlight coming from the host galaxy is responsible for the H{\sc{ii}} classification. Between about 15 and 20\% of the cases exhibit composite emission, and 20\% of the sources are LINER. The presence of LINER-like emission is unexpected but not impossible. Nuclear LINER emission can be produced by an AGN, but extra nuclear LINER-like emission can be explained by fast shocks \citep{1995ApJ...455..468D} or photoionization by late-type stars.

One additional particular case, which we should stress is the example of a galaxy
that can be clearly discriminated in the diagnostic diagram of 6dFGS due to its
extreme location (see Figure \ref{fig:6dfnii}); the galaxy with id 25 (HE 0227-0913) is located in the lower left part of 6dFGS classification
diagram, but unfortunately the lack of the HES data do not allow us to compare
the two classifcations. Previous studies by \cite{2001ApJS..136...61S} found that the galaxy has strong Balmer lines and the ratio
[O\,{\sc{iii}}]/H$\beta$ is uncharacteristically low for the narrow line region
of a Seyfert galaxy, which agrees with 6dFGS results.

\section{Discussion}\label{discussion}

The spectroscopic analysis of the LLQSOs sample helped us examine the properties of these systems. Notably, we  probed the widths of the emission lines and the activity schemes of the samples. Moreover, we  compared these features using two observational methods (fiber and typical long-slit spectroscopy). 

Our targets comprise members of the Seyfert 1 category showing a typical range of emission line widths, FWHM $>$ 1000 km s$^{-1}$ and the widths of the H$\alpha$ and H$\beta$ broad components follow a linear correlation in agreement to \cite{Zhou2006}.

The physical interpretation of
these additional components is not straightforward. 
For example, HE~0236-3101 presents double broad components,while earlier studies have shown that this galaxy is an accretion disk candidate among luminous galaxies \citep{1991ApJ...382L..63C}.  
These broad Balmer emission line profiles are consistent with models of
inclined small relativistic accretion disks around a massive black hole
\citep{1988MNRAS.230..353P, 1989ASNYN...3...19C, 1990ApJ...365L..51H}.

Previous X-ray studies of the double-peaked emitters (Balmer lines) such as
HE~0203-0031 show that the illumination of the accretion disk requires an
external power to produce lines of this strength due to insufficient local power
\citep{2006ApJ...651..749S}. 
However, the fact that the X-ray emission of double-peaked emitters as a class
does not differ from that of normal AGN with similar properties suggests that a
peculiarity of the X-ray emission structure and mechanism is not responsible for
the occurrence of double-peaked Balmer lines in AGN. On the other hand, the
presence of double narrow components may be an indicator for superwinds. 
Galaxies in the local universe with large IR luminosties (L$_{IR}$ $>$ 10$_{44}$ erg/s), large IR excesses (L$_{IR}$ / L$_{OPT}$ $>$ 2), and/or warm far-IR colors (flux density at 60 micron greater than 50\% of the flux density at 100 micron) drive superwinds \citep{ham}. However, the  lack of relevant data and the low spectral resolution prevents us from further conclusions.

The fraction of NLS1s in the 6dfGS data presented is about 10$\%$, which is
consistent with previous findings \citep[e.g., ][]{2008RMxAC..32...86K}.
Nevertheless, there is some discussion, whether NLS1s really represent a class of their own \citep[cf., ][and references therein]{mv2013}. 

\begin{figure}[h]
\begin{center}
\includegraphics[width=\columnwidth]{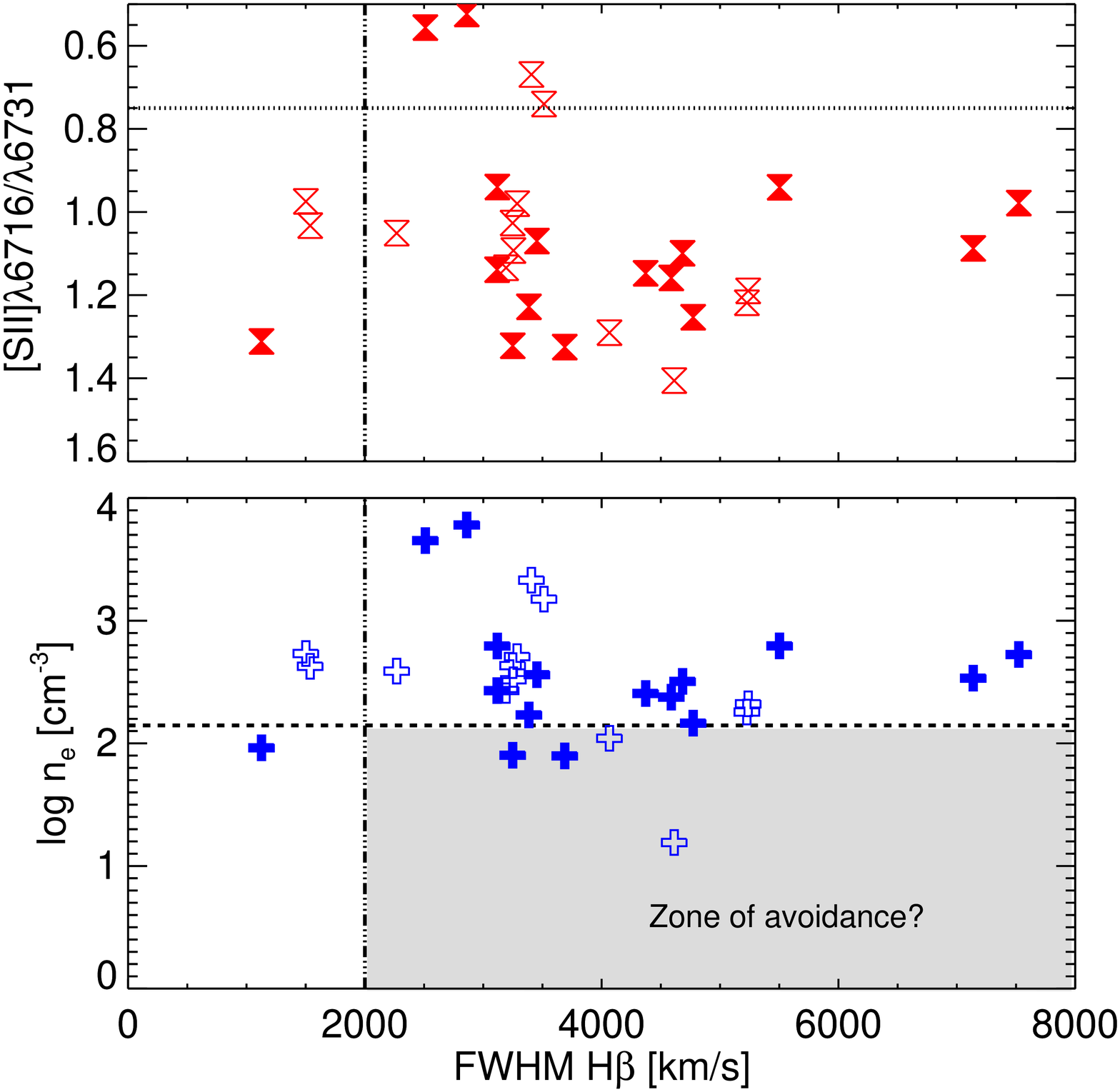}
\end{center}
\caption{FWHM of the H$\beta$ broad component vs. [S{\sc{ii}}]$\lambda$6716/$\lambda$6731 for our sample as measured with the 6df survey. Filled symbols represent galaxies for which the FWHM restriction considered between the broad components of H$\alpha$ and H$\beta$ is fulfilled. The shaded area marks the zone of avoidance for AGNs with broad lines (FWHM(H$\beta$) $>$ 2000 km s$^{-1}$)}, as defined by \citet{Xu2007}.
\label{fig:ne_avoidance}
\end{figure}

\cite{Xu2007} introduced the ``zone of avoidance", where AGNs with lines FWHM H$\beta$ $>$ 2000km s$^{-1}$ avoid low densities, whereas NLSy1 galaxies show a wider distribution in the NLR density, including a number of objects with low densities. 
Outflows may play a key role in driving differences in the NLR between NLS1 and BLS1 (Broad Line Seyfert 1), and consequently the zone of avoidance can be explained \citep{Xu2007}. Additionally, they discuss a number of explanations, such as supersolar metalicities, temperature effects, starburst contributions in NLS1, and the effect of NLR extent, which cannot support the idea of zone of avoidance in density.
The NLR electron density values are consistent with those typically measured in our sample (see Fig\,\ref{fig:ne_avoidance}), although some sources are located in the zone of avoidance. Starburst contribution in a fraction of our sample can lead to lower measured density due to lower average density of H{\sc{ii}} regions. Starburst activity is boosted  in the young objects, since they are still in the process of growing their black holes \citep{mathur}. Moreover, the geometry of the NLR has been suggested to be responsible for the lower average densities, since the density declines
outward in the extended NLR, .

The fraction of LINERs (20\%) was not expected in our sample but their presence can be explained. The [S\,{\sc{ii}}]/H$\alpha$ and the [O\,{\sc{i}}]/H$\alpha$
diagram are considered to be more effective in separating Seyfert from LINER
galaxies \citep{kewley2006}.
The SDSS galaxy sample \citep{kewley2006} shows that the galaxies that are
classified as composite using the [N\,{\sc{ii}}]/H$\alpha$ diagram are mostly
located within the star-forming sequence on the [S\,{\sc{ii}}]/H$\alpha$
diagram. This tendency seems to be followed by our sample as well.
The [O\,{\sc{i}}]/H$\alpha$  diagram is more sensitive to shocks and therefore is a
more reliable tracer of changes in the ionization conditions due to fast shocks.
The other diagrams are more sensitive to star formation. 
Star formation activity can be distributed all over the disk and can produce different ionizing continua.
Additionally, the trend in late-type galaxies studied by \cite{2014MNRAS.441.2296M} agrees with our LINER fraction. Spiral galaxies show lower [O\,{\sc{iii}}] values with increasing aperture. Hence, the unpredictable LINER activity in LLQSOs sample could also be justified as a result of aperture efffect.

\begin{figure*}[!t]
\begin{center}
\includegraphics[width=1.\linewidth]{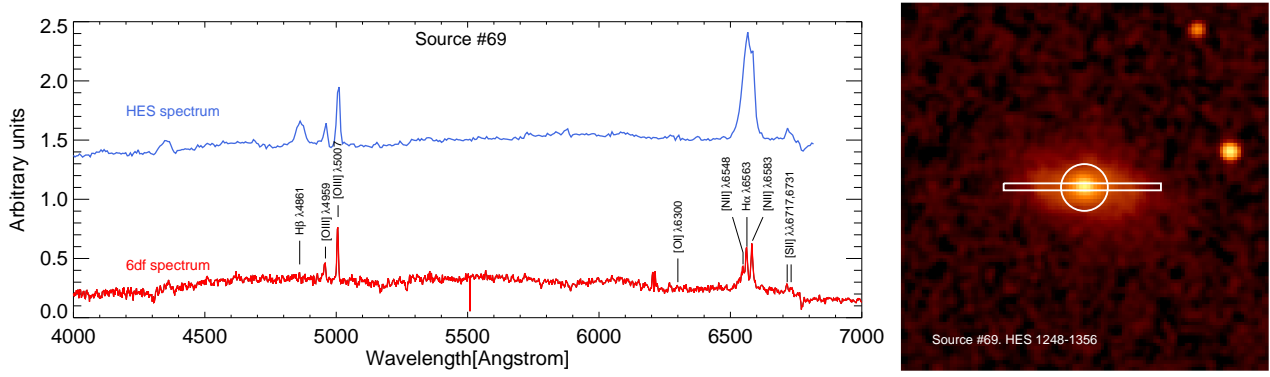}
\end{center}
\caption{Left: top and bottom spectra show the HES and 6df spectra for the same source (\#69, HES1248-1356, redshift 0.0145), respectively. Optical emission lines of interest are labeled. Right: J-H-K$_{s}$ composite 2MASS image of HES1248-1356. White straight lines depict a slit of 2\farcs0 width with EW orientation, as a reference for the slit used in the HES. The white circle represents a 6\farcs7 diameter fiber, as that used in 6dfGS.}
\label{fig:spectral_comparison}
\end{figure*}

\subsection{Aperture effect vs. sources variability}

One unique aspect of the present study is the comparison between two data sets that have been taken with considerably different apertures. When comparing the data sets it is challenging to distinguish between differences that are due to variablility, either in the BLR, NLR or outer regions, or to the aperture effect. 
This would be better addressed using deconvolution techniques to separate the central point source from the background$/$host emission (see, e.g., \citet{Lucy2003}  for successful deconvolution of QSO long-slit data). However, these methods are not applicable in this study. In the case of 6dFGS, each galaxy is observed using a fiber, a single 1D spectra is obtained and hence there is not spatial information. As for the HES, the already extracted 1D spectra we have access to exhibit a variety of SNR in the studied emission lines. H$\alpha$ and [O{\sc{iii}}]$\lambda$5007 are usually strong with a SNR of the order of 40 or more, depending on the case.  As for H$\beta$ its strength varies from about 3 to more than 10. These observations used slits of between  1".5 and 2".5 width. Assuming the extracted galaxy spectra was distributed over a number of N pixels across the slit, this would reduce the observed SNR by a factor of $\sqrt{N}$. In many cases this would render the spatial signal of the H$\beta$ line too weak to be useful to perform a deconvolution. In any case, to apply spectral deconvoution techniques (at least in our long-slit data) we would need the 2D long-slit information and ideally (although not necessarily) a neighboring star to model the PSF. The present discussion is meant to clearly point out the issues faced when comparing different spectral data sets, and to clearly constrain the possible nature of the differences found and exploit the data we have in the best possible way.

\begin{figure*}
\centering \subfloat[]{\label{fig:cross_large}\includegraphics[width=0.45\textwidth]{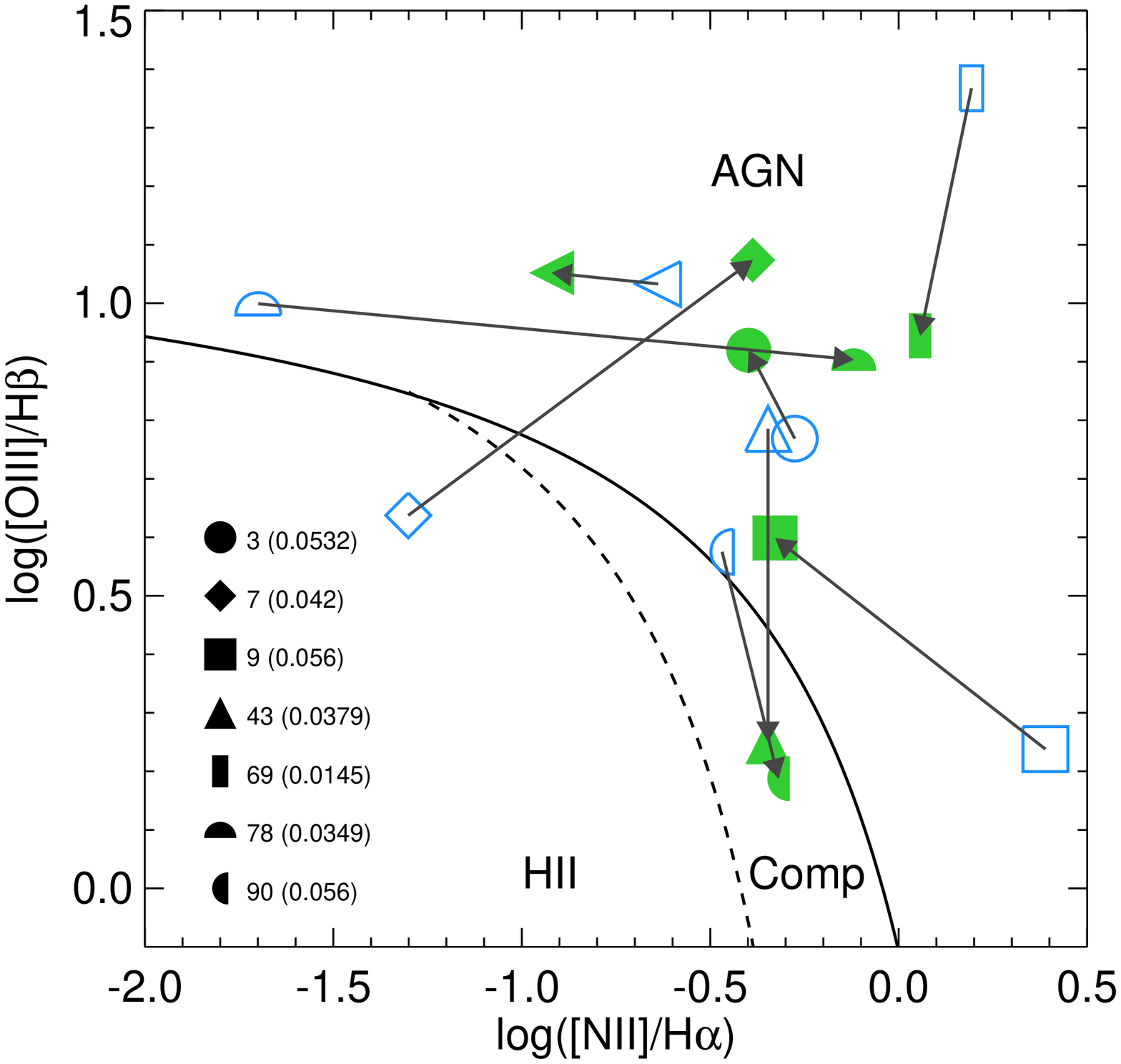}}
\subfloat[]{\label{fig:cross_short}\includegraphics[width=0.45\textwidth]{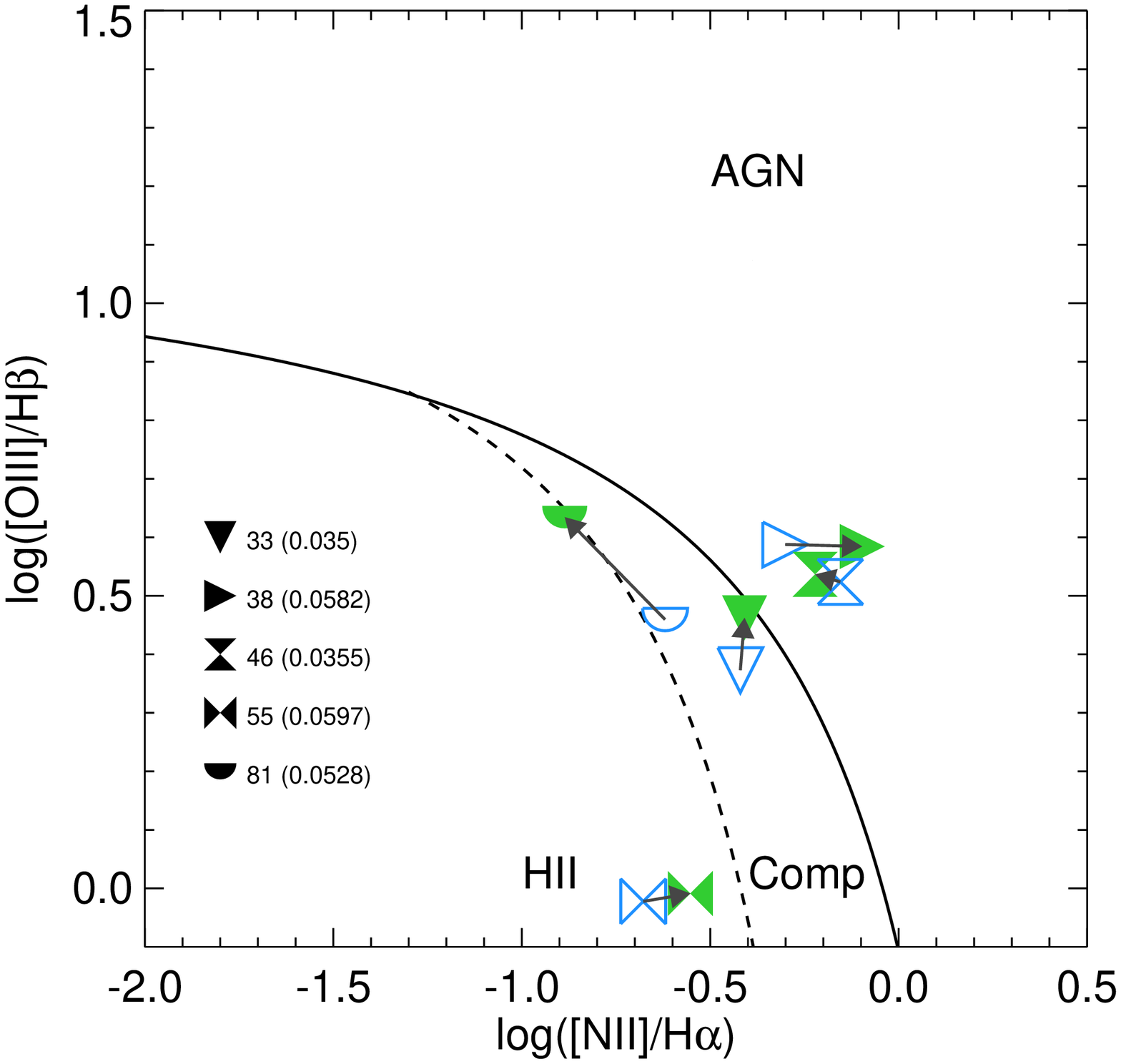}}
\caption[FWHM broad components for H$\alpha$ and H$\beta$]
{\label{fig:cross_both} Diagnostic diagram of cross-matching sources, illustrating the aperture effect. Filled green symbols represent the 6dfGS classification, whereas empty blue symbols show the classification as estimated with the HES. Source identifiers are also given. The 6dfGS data were all taken later in time than the HES, changes in classification are also indicated by  arrows. For all sources, the redshift and years between observations is given. Sources comply with the quality criteria defined for the width of the narrow hydrogen recombination lines in Section\,\ref{emibroad}.}
\end{figure*}

When comparing the HES and 6sFGS spectra, several of our sources presented remarkable differences (see  example in  left panel of Fig.\,\ref{fig:spectral_comparison})  in the relative strength on the emission lines and in the detection as well (e.g., H$\beta$ broad component).
After carefully cross-matching the sources and applying the narrow line widths quality criteria (as defined in Section\,\ref{anal}), we ended up comparing 12 LLQSOs. For some of these sources the classification in the BPT diagram significantly changes depending on the data used (see Fig.\,\ref{fig:cross_both} Left), whereas for others the changes are not significant (see Fig.\,\ref{fig:cross_both} Right). We estimated the average travel distance for these sources to be $\sim$ 0.4 dex in both axes of the BPT diagrams (three times more than the maximum estimated uncertainty of 0.13 dex derived from not correcting the Fe{\sc{ii}} contribution; see section \ref{anal}).

From purely the instrumental point of view, there are two main reasons why a source observed with two different instruments can render different classifications. First, the differences in instrumentation setup; and second, differences in the spectral resolution (see discussion in Sect.\,\ref{emibroad}). In the HES case, the slit width varied between $1\farcs 5$ and $2\farcs 5$ and was placed using an east-west slit position angle, while the 6dfGS uses a 6\farcs7 circular fiber centered in the source. As illustrated in the right panel of Fig.\,\ref{fig:spectral_comparison}, this means that different regions of the galaxy are being observed, which  could easily lead to differences in the spectra and in classification.
In the context of the aperture effect, two main effects can be described: dilution and inclusion \citep{storchi}. 
\begin{itemize}
\item Inclusion of new emitting regions: regions of star formation or ionization cones included in one of the apertures used and rejected by the other could easily be responsible for a change in the source classification. Inclusion effects could change a source classification from H\,{\sc{ii}} to Composite or AGN, or vice-versa.
\item Dilution of the emitting region: Dilution effects can be responsible for an AGN classification moving from higher to lower excitation regions, or simply to composite/H\,{\sc{ii}}. In this context, large amounts of starlight could simply enhance the H\,{\sc{ii}} contribution to the source.
\end{itemize}
These effects are difficult to categorize in an absolute way, as the results really depend on the galaxy geometry and internal structure. However, cases in which the classification of a source is almost the same (as in Fig.\,\ref{fig:cross_both} Right) point toward the source being dominated by its nuclear and close circumnuclear emission.

There is, however, the possibility of sources being variable.
Here two different scenarios are plausible: variability of the emission coming from the narrow lines (e.g., NLR and stellar light) and variability in the BLR. The latter would not have a direct impact in the BPT classification, which uses only the narrow lines, but may affect the fits and dilute the stellar component.
The lifetime of an H\,{\sc{ii}} region is of the order of 10$^{6}$ years, in that period of time there are several events that, if observed, would change a galaxy classification. The bright Helium pulse flashes that occur periodically in AGB stars last for about hundreds years every 10$^{3}\sim$ 10$^5$ years, depending on the stellar mass. These events, even if they are coming from a stellar cluster, are not measurable in the timescales of our observations (ranging from 2 to 12 years). Supernovae explosions can be very bright events, of about M$_B$=-15 to -19 \citep{Richardson2002}, or 10$^{10}$L$\sun$, which would make them detectable in some of our sample cases (see Fig.\,\ref{fig:magn}). They, however, decay very rapidly in timescales of the order of 100-200 days. This makes them difficult to observe in extragalactic observations, but that possibility is open. Finally, the NLR and BLR could potentially undergo variability. The NLR densities are much smaller than in the BLR (n$_{e}$ of 10$^8$ cm$^{-3}$ vs. a maximum of about 10$^{4}$ cm$^-3$ for our LLQSOs, see Fig.\,\ref{fig:electron_density_hist}). Variability in the BLR goes much faster than in the NLR, and in the NLR it lasts  longer, as  the lower density changes do not propagate as fast. So in the timescales of our observations (2-12 years) a change in the BLR could be measured, but not in the NLR.

\begin{figure}[h]
\begin{center}
\includegraphics[width=\columnwidth]{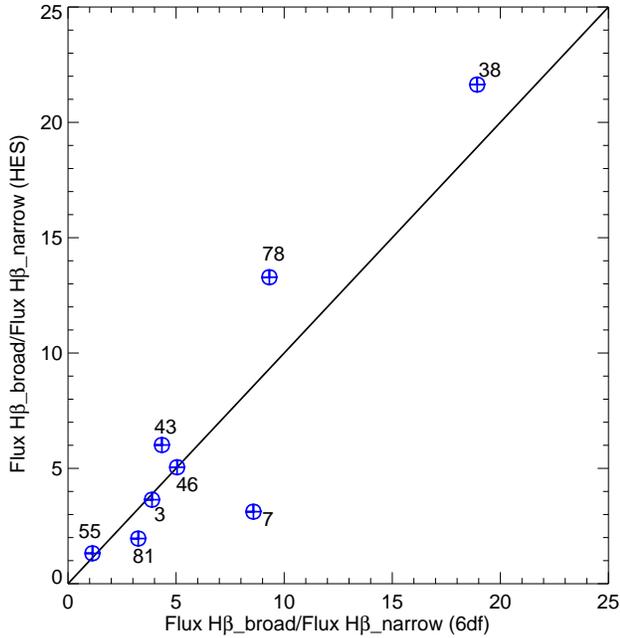}
\end{center}
\caption{Ratio of the H$\beta$ broad component over H$\beta$ narrow component for the cross-matched sources of both samples, which follow the broad lines width quality criteria defined in Section 5.1.}
\label{fig:broad-vs_narrow}
\end{figure}

One way of verifying if the BLR has not undergone variability is to represent the ratio between the narrow and broad line components of the H$\beta$ line (see Fig.\,\ref{fig:broad-vs_narrow}). We chose to use H$\beta$ for this because the spectral resolution, and hence the line fits, are better in that region. For three sources, the relation between the broad and narrow line components remains constant between the years of the HES and 6dfGS observations. This proves the consistency between the convolution and flux conservation of both data sets. Two of them (\#55 and \#46) essentially show no variation in classification based in the narrow lines (see left panel of Fig.\,\ref{fig:cross_both} ), meaning that the sources are nuclear dominated and have not undergone any variability. The third source (\#3) with a constant H$\beta$ line ratio in time shows a relative shift on the activity classification and interestingly located in the zone of avoidance with lower averaged NLR density.
Source \#38 has a constant classification based on the BPT, but its H$\beta$ broad-to-narrow relation changes, making it a candidate for BLR variability. Finally, sources \#7, \#81, and \#43, as well as the rest of the studied LLQSOs, could be suffering from aperture effect as well as from variability. We do not have enough information from the data at hand to disentangle this effects.

\begin{figure}[h]
\begin{center}
\includegraphics[width=\columnwidth]{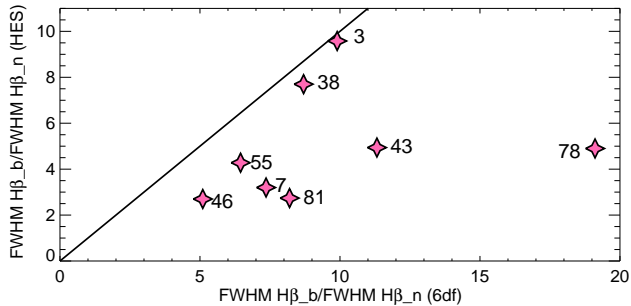}
\end{center}
\caption{Ratio of the FWHM of the H$\beta$ broad component over FWHM of the H$\beta$ narrow component for the cross-matched sources of both samples, which follow the broad lines width quality criteria defined in Section 5.1.}
\label{fig:broad-vs_narrow_fwhm}
\end{figure}

We have also tested the ratio of the broad over narrow H$\beta$ components for the cross-matching sources. As mentioned in Section \ref{emibroad}, our HES sources have not been corrected for instrumental broadening. By using FWHM ratios, we are eliminating any bias that would show up by comparing just measured widths. Figure \ref{fig:broad-vs_narrow_fwhm} shows that the importance of the broad over the narrow component  is systematically higher for our 6dFGS cross-matched sources. This indicates that the better resolution of the data have a direct impact in the results, and that the HES results are clearly biased.

\section{Summary}\label{concl}
We  presented an optical spectroscopy study of a sample of 99 LLQSOs, using two different sets of data. The HES data, observed with a slit and low spectral resolution, and the public 6dFGS data, whose spectra were obtained using a fiber and medium spectral resolution. The aims of the present investigation were to characterize the sample optical spectral properties, classify their activity, and study the effect of changes in aperture vs. variability of the BLR and$/$or NLR.
We found the narrow and broad line widths to be within the expected values, on average several thousand km$/$s for the broad and $\lesssim$1000\,km$/$s for the narrow components. However, the results were sensitive to the spectral resolution and continuum characteristics of the spectra. For this reason, several fitting constrains and post-processing quality flags were used to ensure the consistency of the results. A small number of galaxies presented no broad component, but it is likely that they were buried in the data noise. We also found two sources with spectral characteristics consistent with the presence of double broad components, and six galaxies that comply with the classic NLS1 requirements.
As for the NLR electron density value of our sources, it exhibits lower densities $\sim$ 2 to 3 cm$^{-3}$, consistent with NLR regions of broad line AGNs. The results are  in good agreement with the zone of avoidance presented by \cite{Xu2007}.
We also tested the relevance of Population A vs. Population B. Using only optical-based indicators, we find that 50\% of our sources with H$\beta$ broad emission fit well into the \citet{Sulentic2000, Sulentic2000b} radio-quiet sources definition. The remaining sources could be interpreted as low-luminosity, radio-loud quasars.

We presented the activity classification schemes (BPT diagrams)  for both surveys, showing an AGN/Seyfert activity of between 50$-$60\%. 
The starburst contribution throughout the galaxy might control the LINER and H\,{\sc{ii}} classification of our sample. 
We justified the notable differences in classification from the comparison of the HES and 6dfGS spectra in the context of the aperture effect. It is challenging to quantify these effects  since they depend on the galaxy structure and geometry. Alternatively, the differences measured could be due to variability in the BLR, although the nature of our data prevents us giving from a  more precise quantification.

\begin{acknowledgements}
The HES data was kindly provided by Lutz Wisotzki, who also participated in discussions and gave extensive 
contributions to the paper. Special thanks to Marios Karouzos for very useful suggestions and discussions.
This work is partially a result of the collaborative project between Korea Astronomy and Space Science Institute and Yonsei University through DRC program of Korea Research Council of Fundamental Science and Technology(DRC-12-2-KASI). This work has also been supported by the National Research Foundation of Korea grant 2012-8-0582.
Part of this work was supported by the German {\it Deut\-sche
For\-schungs\-ge\-mein\-schaft, DFG\/} project numbers SFB956 and SFB494. Macarena Garcia-Marin is supported by the German federal department for education and research (BMBF) under the project number 50OS1101.
Mariangela Vitale is member of the International Max-Planck Research School (IMPRS) for 
Astronomy and Astrophysics at the Universities of Bonn and Cologne, supported by
the Max Planck Society. JZ acknowledges support by the German Academic Exchange
Service (DAAD) under project number 50753527. This project has made use of the
Final Release of 6dFGS data \citep{jones2004, jones2009}. 
This research has also made use of the NASA/IPAC Extragalactic Database (NED)
which is operated by the Jet Propulsion Laboratory, California Institute of
Technology, under contract with the National Aeronautics and Space
Administration. While working on the current research, Ned's Wright cosmology
calculator was also used \citep{2006PASP..118.1711W}. 
\end{acknowledgements}

\bibliography{llqso}

\bibliographystyle{aa} 
\onecolumn
\begin{landscape}\tiny

\LTcapwidth=\textwidth
        \centering
\begin{longtable}[width=\textwidth]{cccccccccc}
\caption{\label{tab:6df}Fitting results of 6dFGS survey. The luminosities
ratios and and both narrow and broad FWHM of the Hydrogen lines are shown. An
additional column with some comments describes broadly the quality of the
spectra and the emission of the lines. $\ddagger$ and $\dagger$ mean that the source complains with the narrow and broad line widths requirements defined in Section\,\ref{emibroad}, respectively.
}\\   
\hline\hline
\rule[-1ex]{0pt}{1.5ex}  Comments & ID & [N\,{\sc{ii}}]/H$\alpha$ & [O\,{\sc{iii}}]/H$\beta$ & [S\,{\sc{ii}}]/H$\alpha$ & [O\,{\sc{i}}]/H$\alpha$ &\multicolumn{2}{c}        { FWHM H$\alpha$ (km/s)}&\multicolumn{2}{c} {FWHM H$\beta$ (km/s)}\\
& &  & & & &Narrow Component& Broad Component& Narrow Component& Broad Component\\
\hline
\endfirsthead
\caption{continued}\\
\hline\hline
\rule[-1ex]{0pt}{1.5ex}  Comments & ID & [N\,{\sc{ii}}]/H$\alpha$ & [O\,{\sc{iii}}]/H$\beta$ & [S\,{\sc{ii}}]/H$\alpha$ & [O\,{\sc{i}}]/H$\alpha$ &\multicolumn{2}{c}        { FWHM H$\alpha$ (km/s)}&\multicolumn{2}{c} {FWHM H$\beta$ (km/s)}\\
& &  & & & &Narrow Component& Broad Component& Narrow Component& Broad Component\\
\hline
\endhead
\hline
\endfoot
$\ddagger$        &       3   &      0.40    &    8.31    &     0.66 &    3.45  &  423.5   &     2844.6    & 465.8    &  4612.9 \\                 
No [O\,{\sc{i}}]$\ddagger$$\dagger$  &       4   &      0.23    &    6.01    &     0.37 &    ---  &  585.2   &     2669.4    & 472.8    &  3385.4  \\
NLS1 $\ddagger$   &       5  &      0.26    &    1.46    &     0.15 &    0.02  &  675.3   &     2382.9    & 608.0    &  1537.8 \\ 
$\ddagger$$\dagger$&       7   &      0.41    &    11.85   &    1.49 &    0.50  &  345.5   &     2182.2    & 423.4    &  3116.7 \\
NLS1$\ddagger$              &       8   &      0.78    &    0.23    &     0.31 &    0.03  &  326.1   &     1246.9    & 335.7    &   526.2 \\
No H$\beta$ broad $\ddagger$ &       9   &      0.47    &    3.97    &     0.77 &    0.21  &  496.5   &     2813.1    & 408.7    &  ---    \\
$\ddagger$                  &       12  &      0.45    &    1.78    &     0.44 &    0.04  &  324.9   &     1488.9    & 403.4    &  3248.1 \\
{\bf N.B.E.}      &       13  &      0.70    &    3.32    &     0.48 &    0.12   &  409.1   &      328.8    &  ---     &  ---    \\
out               &       14  &      ---     &    ---     &     ---  &    ---   &  ---     &     ---       &  ---     &  ---    \\
No H$\beta$ broad $\ddagger$ &       19  &      0.85    &    7.80    &     0.43 &    0.22  &  381.7   &     33756.0   &  439.0   &  ---    \\
No H$\beta$ broad &       20  &      0.48    &    6.97    &     0.93 &    3.64  &  311.1   &     4241.7    &  438.9   &  ---    \\
No [O\,{\sc{i}}], Uncertain [N\,{\sc{ii}}]  &       22  &      ---   &    13.04   &     0.02 &    ---   &  377.8   &     6275.2    &  532.4   &  3512.8  \\
$\ddagger$        &       23  &      1.34    &    24.99   &     0.53 &    0.32  &  502.3   &     3072.6    &  561.1   &  ---    \\
$\ddagger$$\dagger$ &       24  &      0.47    &    8.88    &     0.84 &    0.24  &  481.9   &     3709.1    &  477.5   &  4371.6 \\
No [O\,{\sc{i}}] $\ddagger$        &       25  &      0.04    &    0.39    &  0.82 &    ---  &  546.7   &     1937.2    &  718.3   &  3254.4 \\
No [O\,{\sc{i}}] $\ddagger$$\dagger$  &       26  &      0.48    &    11.32   &     0.72 &    ---   &  558.8   &     4791.1    &  558.0   &  5502.8  \\
NLS1 $\dagger$    &       29  &      0.91    &    10.49   &    0.78 &    0.18 &  410.6   &     1803.9    &  622.2   &  1848.8 \\          
$\dagger$         &       30  &      0.62 hhhh   &    3.49    &     0.36 &    0.17  &  492.5   &     3230.4    &  392.3   &  3453.0 \\
No H$\beta$ broad &       32  &      1.52    &    2.87    &     0.26 &    0.10  &  652.2   &     3982.4    &  484.3   &  ---    \\
$\ddagger$        &       33  &      0.39    &    2.90    &     0.68 &    0.14  &  384.5   &     2183.9    &  407.4   &  4065.7 \\
$\ddagger$        &       35  &      0.20    &    5.57    &    1.27&    1.07  &  433.4   &     3214.3    &  466.8   &  5237.6 \\
                  &       36  &      5.26    &    3.15    &    1.53 &    7.30  &  597.7   &     4540.5    &  285.2   & 14403.6 \\      
$\ddagger$        &       38  &      0.80    &    3.84    &     0.40 &    0.16  &  475.2   &     5248.5    &  537.3   &  4705.2 \\ 
$\ddagger$$\dagger$ &     40  &      0.72    &    6.59    &     0.25 &    0.20  &  393.7   &     2990.3    &  476.5   &  4172.0 \\
$\ddagger$$\dagger$ &     41  &      0.45    &    1.69    &     0.28 &    0.12 &  391.9   &     3162.2    &  481.9   &  4683.6 \\
                  &       42  &      0.57    &    10.24   &    0.88 &    0.79  &  402.9   &     9959.6    &  541.7   &  5226.5 \\ 
$\ddagger$$\dagger$ &     43  &      0.12    &    11.27   &    0.91 &    0.36  &  448.8   &     3489.2    &  421.6   &  4773.0 \\
$\ddagger$$\dagger$ &     44  &      0.25    &     6.49   &    0.37 &    0.06  &  494.2   &     2433.9    &  422.8   &  3320.1 \\
No [O\,{\sc{i}}], [S\,{\sc{ii}}] $\ddagger$$\dagger$ & 46 &0.60    &    3.44    &     0.20  &    ---   &  832.8   &     3159.7    &  899.0   &  4588.0   \\
{\bf N.B.E.} $\ddagger$ & 51  &      0.15    &    2.75    &  1.29 &    0.20  &  332.5   &     ---       &  392.2   &  ---    \\
out               &       52         &      ---     &    ---     &     0.46  &    1.63   &  ---     &     3187.0    &  ---     &  ---    \\
out               &       53         &      ---     &    ---     &     0.90  &    ---  &  ---     &     7369.8?   &  ---     &  ---      \\
NLS1 $\ddagger$$\dagger$ & 54  &      0.20    &    3.52    &     1.34 &    0.28  &  428.1   &     1734.1    &  434.4   &  1886.5 \\
No [O\,{\sc{i}}] $\ddagger$ &   55  &      0.28    &    0.98    &     0.57 &    ---   &  451.7   &     1596.7    &  509.5   &  3286.1  \\
$\ddagger$        &       58  &      0.13    &    2.62    &  0.58 &    0.18  &  558.5   &     1982.9    &  503.3   &  3192.6 \\
No [O\,{\sc{i}}]  &       59  &      0.39    &    3.12    &  0.08 &    ---   &  823.3   &     2829.7    &  557.7   &  2268.6  \\
$\dagger$         &       62 &      0.65    &    2.00    &     0.33 &    0.07  &  408.8   &     2209.0    &  589.4   &  3248.9 \\
$\ddagger$$\dagger$ &     64  &      0.65    &    1.91    &     0.25 &    0.55  &  486.2   &     2352.0    &  479.1   &  3014.8 \\
No H$\beta$ broad &       66  &      0.39    &    8.50    &     0.90 &    0.40  &  596.9   &     3314.9    &  471.8   &  ---    \\
No H$\beta$ broad No [O\,{\sc{i}}] $\ddagger$ &   69  &      1.14    &    ---    &     0.29 &   ---  &  369.0   &     2915.6    &  431.6   &  ---    \\
$\ddagger$$\dagger$ &     71  &      0.13    &    9.42    &     0.95 &    0.73  &  417.5   &     2744.8    &  455.9   &  3688.3 \\
$\dagger$         &       72  &      1.96    &    2.22    &     0.17 &    3.42  &  285.4   &     1922.6    &  1146.3  &  2860.3 \\
                  &       73  &      1.53    &    11.99   &    0.64 &    0.25  &  554.8   &     3768.3    &  743.6   &  7556.9 \\
$\dagger$         &       74  &      0.36    &    2.83    &    0.22 &    1.18 &  626.4   &     2147.3    &  471.6   &  2510.8 \\
$\dagger$         &       76  &      1.02    &    13.39   &     0.89 &    0.37  &  524.1   &     4919.8    &  761.5   &  4811.7 \\ 
NLS1              &       77  &      0.01    &    2.78    &     1.53 &    3.59  &  816.4   &     2625.9    &  398.1   &  1501.5 \\ 
$\ddagger$$\dagger$ &     78  &      0.76    &    8.01    &     0.65 &    0.10  &  368.3   &     5198.1    &  373.4   &  7138.5 \\
$\ddagger$        &       81  &      0.13    &    4.31    &     1.34 &    0.65  &  405.4   &     1741.4    &  414.1   &  3406.6 \\
No H$\beta$       &       82  &      ---     &    9.28    &     --- &    ---  &  369.1   &     4515.9    &  ---     &  ---    \\
No [O\,{\sc{iii}}]&       83  &      1.00    &    ---     &     0.63 &    0.24  &  509.8   &     4742.2    &  392.0   &  ---    \\
$\ddagger$$\dagger$ &     84  &      0.95    &    7.56    &     0.24 &    0.15  &  476.9   &     5253.2    &  382.9   &  7523.7 \\
out               &       86         &      ---     &    ---     &     ---  &    0.29   &  ---     &     3174.9    &  ---     &  ---    \\
$\ddagger$$\dagger$ &     87  &      0.13    &    8.69    &     1.10 &    1.00  &  446.4   &     2620.8    &  412.4   &  3118.6 \\
No H$\beta$ broad, No [O\,{\sc{i}}] $\ddagger$ &  90  &      0.48    &    1.54    &     1.30 &   ---  &  385.4   &     2647.9    &  342.0   &  ---    \\
No [O\,{\sc{i}}], NLS1 $\dagger$&   91  &      0.28    &    0.72    &    0.16 &    ---   &  856.7   &     1124.4    &  ---     &  1126.7  \\               
\bf{N.B.E.} $\ddagger$$\dagger$  &       94  &      0.59    &    2.96    &     0.50 &    0.16  &  623.1   &     ---       &  549.3   &  ---    \\
$\ddagger$$\dagger$ &     96  &      1.22    &    13.07   &     0.10 &    0.17  &  700.0   &     1462.7    &  628.4   &  2204.3 \\
                  &       97  &      1.14    &    17.16   &     0.52 &    0.59  &  330.2   &     4815.9    &  457.2   & 11037.7 \\      
\hline
        
\end{longtable}
\end{landscape}


\begin{landscape}\tiny
\LTcapwidth=\textwidth
\centering

\begin{longtable}[width=\textwidth]{cccccccc}
\caption{\label{tab:hes}Fitting results of HES survey. The luminosity ratios
and both narrow and broad FWHM of the Hydrogen lines are shown. An
additional column with some comments describes broadly the quality of the
spectra and the emission of the lines. $\ddagger$ and $\dagger$ mean that the source complains with the narrow and broad line widths requirements defined in Section\,\ref{emibroad}, respectively.}\\
\hline\hline
\rule[-1ex]{0pt}{1.5ex}Comments &  ID & [N\,{\sc{ii}}]/H$\alpha$ & [O\,{\sc{iii}}]/H$\beta$ & \multicolumn{2}{c}{ FWHM H$\alpha$ (km/s)} & \multicolumn{2}{c} {FWHM H$\beta$ (km/s)}\\
&  & & &Narrow Component& Broad Component& Narrow Component& Broad Component\\
\hline
\endfirsthead
\caption{continued}\\
\hline\hline
\rule[-1ex]{0pt}{1.5ex}Comments &  ID & [N\,{\sc{ii}}]/H$\alpha$ & [O\,{\sc{iii}}]/H$\beta$ & \multicolumn{2}{c}{ FWHM H$\alpha$ (km/s)} & \multicolumn{2}{c} {FWHM H$\beta$ (km/s)}\\
&  & & &Narrow Component& Broad Component& Narrow Component& Broad Component\\
\hline
\endhead
\hline
\endfoot
                     & 1     &   0.23  & 3.23     &  1127         &    3396       &  672          &  2264 \\ 
No H$\beta$ broad    & 2     &   1.08  & 8.24     &  479          &    11920      &  746          &  --- \\
$\ddagger$$\dagger$  & 3     &   0.53  & 5.87     &  777          &    5460       &  667          &  6394 \\
$\ddagger$$\dagger$  & 6     &   0.52  & 3.42     &  842          &    4590       &  999          &  5293  \\
$\ddagger$           &7   &   0.05  & 4.34     &  921          &     1432      &  1151         &  3647  \\
No H$\beta$ broad    & 8     &   0.89  & 1.89     &  932          &    1474       &  1561         &  ---\\    
$\ddagger$           & 9    &   2.45  & 1.73     &  1550         &    1545       &  1377         &  4162 \\ 
$\dagger$            & 11    &   0.84  &  5.86    &  931          &    3201       &  1379         &  3645  \\ 
$\ddagger$$\dagger$  & 12    &   0.45  &  6.10    &  1082         &    6440       &  1115         &  7589  \\ 
$\dagger$            & 15    &   0.43  &  1.09    &   785         &    3784       &  1114         &  4392 \\ 
$\ddagger$$\dagger$  & 16    &   0.76  &  8.53    &  1201         &    4559       &  1489         &  5299 \\ 
$\ddagger$$\dagger$  & 17    &   0.29  &  1.17    &  1391         &    5485       &  1674         &  7038 \\ 
$\ddagger$$\dagger$  & 18    &   0.17  &  4.95    &  1667         &    4445       &  1600         &  5186  \\ 
$\dagger$            & 19    &   0.70  &  7.80    &   866         &    4612       &  1234         &  7079   \\ 
$\ddagger$$\dagger$  & 20    &   1.72  &  18.21   &   666         &    4839       &  602          &  6178 \\ 
No H$\alpha$ broad   & 21    &   0.26  &  1.46    &   802         &    ----       &  1540      &  1495?? \\ 
$\dagger$            & 22    &   0.15  &  12.9    &  2217         &    7737       &  789          &  7502 \\ 
                     & 23    &   0.20  &  3.30    &  2067         &    7451       &  967          &  5115 \\ 
No H$\alpha$ fitted  & 24    &   ---   &  8.51    &  ---          &    ---        &  600          &  5653 \\ 
                     & 27    &   0.65  &  1.49    &  831          &    4486       &  1633         &  3879(blue)/3400.94(blue) \\ 
No H$\alpha$ fitted  & 28    &   ---   &  16.25   &  ---          &    ---        &  1113         &  6043(blue)/5220(red) \\ 
                     & 29    &   0.88  &  3.81    &  671          &    2172       &  1329         &  4079 \\ 
                     & 30    &   0.73  &  2.38    &  1036         &    3346       &  1630         &  6834  \\ 
$\ddagger$$\dagger$  & 31    &   0.54  &  4.91    &   977         &    4362       &   848         &  5660 \\ 
No H$\beta$ broad    & 32    &   1.39  &  2.62    &   855         &    3672       &  1148         &  ----\\ 
$\ddagger$$\dagger$  & 33    &   0.38  &  2.36    &  1226         &    3678       &  1186         & 4544  \\ 
$\ddagger$           & 34    &   0.61  &  4.58    &   777         &    4396       &  1015         & 7184  \\ 
$\ddagger$$\dagger$  & 37    &   0.10  &  0.54    &   928         &    4021       &  1223         & 6046  \\
$\ddagger$$\dagger$  & 38    &   0.50  &  3.87    &   669         &    3900       &   677         & 5224  \\
$\dagger$            & 39    &   0.11  &  5.98    &  1704         &    3616       &  1187         & 3462  \\
$\dagger$            & 40    &   0.69  &  13.98   &   698         &    4208       &   525         & 4059  \\
No H$\beta$ broad    & 41    &   0.29  &   0.24   &   765         &    2414       &  1405         & ----  \\
$\dagger$            & 42    &   1.99  &   9.48   &  1576         &    8127       &  1066         & 9432  \\
$\ddagger$$\dagger$  & 43    &   0.23  &  10.79   &  1039         &    4870       &  1203         & 5948  \\
$\dagger$            & 44    &   0.99  &   1.16   &  1142         &    4541       &  1609         & 6897  \\
$\ddagger$           & 45    &0.19&   2.36   &584   &4367  &   568         & 1846  \\
$\ddagger$$\dagger$  & 46    &   0.70  &   3.34   &  999          &    3371       &  1209         & 3628  \\
                     & 47    & 0.39    &   3.96   &  1871         &    4589       &  1203         & 4087  \\
$\dagger$            & 48    & 1.85    &   1.90   &   602         &    4487       &  1317         & 6247  \\
$\ddagger$$\dagger$  & 49    &   0.13  &   1.82   &  1172         &    3906       &  1187         & 4531  \\
\bf{N.B.E.} $\dagger$ & 51   &   0.12  &   1.01   &   993         &     ---       &  2107         & ---   \\ 
$\ddagger$           & 53    &   0.55  &   0.62   &  1008         &    4725       &  1110         & 9256  \\ 
                     & 54    &   0.20  &   1.30   &  1047         &    3230       &  1401         & 5279  \\ 
$\ddagger$           & 55    &   0.21  &   0.95   &   883         &    2878       &  1093         & 4679  \\ 
$\ddagger$$\dagger$  & 57    &   0.24  &   1.98   &  1194         &    3012       &  1589         & 3486  \\
                     & 58& 0.09  &   1.77   &  1264         &    3787       &  1784         & 6782  \\
$\ddagger$$\dagger$  & 59    &   0.25  &   1.58   &  1516         &    3031       &  1279         & 2994  \\
No H$\alpha$ broad   & 60    &   0.25  &   2.59   &  1143         &    ----       &  1606         & 4556  \\
$\dagger$            & 61 & 1.23 &   3.97   &   586         &    4403       &  1280         & 5129  \\
$\dagger$            & 62    &   0.78  &   2.09   &  1067         &    4321       &  1566         & 5043  \\
                     & 63&0.24&0.38&498&5125  &\ ---    & ---     \\
                     & 64&0.63&   1.35   &1675 & 3878 &  1012         & 2356  \\
                     & 65    &   0.10  &   1.42   &  1969         &    6494       &  1466         & 3019  \\
No H$\beta$ broad    & 66    &   0.36  &   9.73   &   400         &    3724       &   771         & ----  \\
$\ddagger$           & 69    &1.56&  23.3    & 890  & 4171 &   849         & 2387  \\
$\dagger$            & 71    &   0.11  &   6.22   &   433         &    3119       &   738         & 4491  \\
$\ddagger$$\dagger$  & 72    &   0.47  &   0.93   &  1285         &    3028       &  1704         & 3498  \\
 out                 & 73    &   ---   &   ---    &  ---          &    ---        &  ---          & ---   \\
$\dagger$            & 74   &   0.07  &   0.79   &  1129         &    4841       &   591         & 6317  \\
                     & 75&0.51&   2.16   &3646  &5551  &  1020         & 4540  \\
$\ddagger$           & 78&0.02&   9.98   &1324  &1608  &  1618         & 7918  \\
$\ddagger$           & 79    & 0.15    &   4.34   &   681         &    3613       &   602         & 6201  \\
                     & 80    & 0.51    &   3.94   &   934         &    3271       &  1339         & 2769  \\
No H$\alpha$ broad $\ddagger$ & 81    & 0.24    &   2.88   &  1179         &    ----       &  1277         & 3507  \\
$\dagger$            & 82    & 1.00    &   16.65  &  2106         &    5046       &  1290         & 5822  \\
$\ddagger$$\dagger$  & 90    & 0.34    &    3.76  &  1699         &    6255       &  1544         & 7234  \\
No H$\beta$ broad    & 91    & 0.57    &    0.48  &   527         &    1727       &  1299         & ----  \\
$\ddagger$           & 92 & 0.53 & 2.74  &  1024         &    3663       &  1325         & 7774  \\
$\ddagger$$\dagger$  & 93    & 1.11    &   10.70  &   670         &    8877       &   891         &11825  \\
$\dagger$            & 97    & 0.62    &    6.39  &   519         &    5877       &  1227         & 7343 \\ 
$\dagger$            & 98    & 0.37    &    6.89  &   500         &    1767       &   764         & 1892 \\ 
$\ddagger$           & 99    & 0.32    &    0.79  &  1184         &    3672       &  1153         & 2625  \\
\hline

\end{longtable}
\end{landscape}


\end{document}